\documentclass[12pt]{article}
\usepackage{color,amsmath,amssymb,exscale,epsfig}
\usepackage{hyperref}
\usepackage{slashed,epsfig,graphicx}

\newcommand{\be}{\begin{equation}}
\newcommand{\bea}{\begin{eqnarray}}
\newcommand{\ee}{\end{equation}}
\newcommand{\eea}{\end{eqnarray}}

\newcommand{\m}[1]{\marginpar{{\tiny *}} }

\def\bea{\begin{eqnarray}}
\def\eea{\end{eqnarray}}

\definecolor{nicered}{rgb}{0.7,0.1,0.1}
\definecolor{nicegreen}{rgb}{0.1,0.5,0.1}
\hypersetup{colorlinks,citecolor= nicegreen,linkcolor= nicered}

\begin{document}
\topmargin -1.0cm
\oddsidemargin -0.9cm
\evensidemargin -0.9cm

  \newcommand{\AddrJSI}{{\sl \small J. Stefan Institute, \\\sl \small
      Jamova 39, P. O. Box 3000, 1001 Ljubljana, Slovenia}}
  \newcommand{\AddrFMF}{{\sl \small Department of Physics, University
      of Ljubljana, \\\sl \small Jadranska 19, 1000 Ljubljana,
      Slovenia}} 
\newcommand{\AddrUBA}{{\sl \small CONICET, IFIBA, Departamento de F\'isica, FCEyN, Universidad de Buenos Aires, \\\sl\small 
	Ciudad Universitaria, Pab. 1, (1428), Ciudad de Buenos Aires, Argentina}}
\newcommand{\AddrUNSL}{{\sl \small CONICET, INFAP, Departamento de F\'isica, FCFMyN, Universidad Nacional de San Luis,  \\\sl\small
	Av. Ej\'ercito de los Andes 950, (5700), San Luis, Argentina}}
 \newcommand{\AddrMUN}{{\sl \small TUM Institute for Advanced Study, \\\sl \small
 Lichtenbergstr. 2a, D-85747 Garching, Germany}}

\hfill \footnotesize{FLAVOUR(267104)-ERC-58}
\vspace{30pt}
\begin{center}
{\Large {\bf  Leptonic Monotops at LHC}}
  \\[15mm]
  Ezequiel Alvarez$^{a,}$\footnote{e-mail address: sequi@df.uba.ar}, 
  Estefania Coluccio Leskow$^{a,}$\footnote{e-mail address: ecoluccio@df.uba.ar}, 
  Jure Drobnak$^{b,c,}$\footnote{e-mail address: jure.drobnak@tum.de},
  Jernej F. Kamenik$^{c,d,}$\footnote{e-mail address: jernej.kamenik@ijs.si} 
  \vspace{1cm}\\
  $^a$\AddrUBA.\vspace{0.4cm}\\
  $^b$\AddrMUN.\vspace{0.4cm}\\
  $^c$\AddrJSI.\vspace{0.4cm}\\
  $^d$\AddrFMF.\vspace{0.4cm}\\
\end{center}

\begin{abstract}

We study the possibility of detecting New Physics (NP) phenomena at the LHC through a new search strategy looking at the monotop (top plus missing energy) signature which is common to a variety of NP models.  We focus on the leptonic top decay mode and study the discovery or exclusion reach of the 2012 LHC data for three example models. Contrary to the hadronic mode, in this case the problematic QCD multijet background can be safely neglected.  We find that the key kinematic variable to suppress most of the remaining SM backgrounds is the transverse mass of the charged lepton and missing energy.  In fact, one could expect that the single-top production measurements already address the monotop signature in this mode.  This is however not the case because in the SM single-top production the transverse mass has an end point determined by the $W$ mass, while the NP signals typically have an additional source of missing energy.  We compare, under the same conditions, our monotop search strategy with existing single-top measurements and find a considerable improvement in the monotop signature reach.

\end{abstract}

\newpage

\section{Introduction}

In the Standard Model (SM) the top quark induces the most severe hierarchy problem. Furthermore, in most natural models it is linked to electroweak symmetry breaking. Consequently, there is strong motivation to search for new physics (NP) effects associated with top physics. 

In fact, possible hints of non-standard contributions in $t\bar t$ pair production have been reported~\cite{Aaltonen:2011kc,CDF:2012xba,Abazov:2011rq}. The inclusive forward-backward asymmetry ($A_{FB}$) in the $t\bar t$ rest frame has been measured by both the CDF~\cite{Aaltonen:2011kc,CDF:2012xba} and D\O~\cite{Abazov:2011rq} collaborations and found to be significantly larger than the SM prediction.  Furthermore, a larger than expected asymmetry measured in bins of the $t\bar t$ invariant mass and $t\bar t$ rapidity difference has also been reported. A related observable at the LHC is the charge asymmetry in $t\bar t$ production, $A_C$. In contrast to the forward-backward asymmetries, the measurements of $A_C$~\cite{ATLAS:2012an,CMS:2012eua} agree with the SM expectations. 

{On the example of a simple $\bar t \slashed{Z'} u$ model with $m_{Z'}> m_t$, it has been shown recently that the simultaneous agreement with the anomalously large $A_{FB}$ and the SM-like $A_C$ measurements, can be achieved provided $\mathcal B(Z' \to u \bar t) \sim (1/3-1/4)$\,  while constraints from $tj$ resonance searches and measured jet multiplicities in inclusive $t\bar t$ production can be simultaneously avoided~\cite{1209.4354,1209.4872}}.\footnote{For an alternative mechanism to simultaneously accommodate $A_{FB}$ and $A_C$ in $s-$channel models see~\cite{Drobnak:2012cz,AguilarSaavedra:2012va}.} The total $Z'$ decay width should thus be dominated by other final states. An intriguing possibility then is that the $Z'$ couples dominantly to a hidden sector resulting in a sizable $\Gamma(Z' \to $invisible). In this case, the suppression of $A_C$ (and contribution to $t\bar t j$ ) can be directly  correlated with the appearance of the monotop ($t+\slashed E_{T} $) signature~\cite{1106.6199,1107.0623}. 
Working with the $Z'$ example, in the present paper we demonstrate how monotops can weight in on the $A_{FB}/A_C$ puzzle in $t\bar t$ production at the Tevatron and LHC.

The monotop signal is also predicted in many other NP settings, most notably in models linking cosmological dark matter (DM) with flavor dynamics~\cite{1107.0623}. On the one hand, the agreement of SM predicted small FCNCs with the precision flavor experiments requires any NP at the TeV scale to have a highly nontrivial flavor structure. Only small amount of flavor violation is allowed phenomenologically.  On the other hand, due to loop and renormalization group effects involving SM Yukawas, some amount of flavor violation in the interactions between DM and SM sector is unavoidable (c.f.~\cite{Kamenik:2011vy}). It turns out that in models where the DM couples to SM quarks via new scalar interactions,  the monotop signal typically dominates over other DM collider signatures like monojets~\cite{1107.0623}. We consider one such example DM model, namely a Two Higgs Doublet Model  (THDM) coupled to a $Z_2$ symmetric neutral scalar -- the DM~\cite{Li:2011ja}. The model has been shown to remain viable in light of recent direct DM detection and invisible Higgs decay width constraints for special values of the scalar potential parameters, provided DM annihilation proceeds predominantly through the heavier of the two CP even neutral Higgses ($H$).  Assuming a natural size of the flavor violating couplings of $H$ we explore the model's signal discovery reach using the monotop signature with existing LHC data. 

Finally, flavor-changing neutral current top quark decays $t \to u(c) Z$ are already the subject of an extensive experimental program at the LHC~\cite{atlasFCNC,cmsFCNCold,cmsFCNC}. They are predicted to be tiny in the SM~\cite{Eilam:1990zc}, while several well motivated NP scenarios predict observable rates (c.f.~\cite{AguilarSaavedra:2004wm} for a review). If mediated by heavy new degrees of freedom, integrated out above the weak scale, the relevant dynamics can be conveniently parametrized in terms of SM gauge invariant effective operators. Encoded in this way,  the rare $t \to u(c) Z$ decays can be linked to several other related processes. In particular, most of the lowest dimensional operators are constrained indirectly by precision $B$ physics observables~\cite{Fox:2007in}.  At the LHC however, one can relate the $t \to u(c) Z$ decay to associated $tZ$ production. Given the sizable $\mathcal B(Z\to {\rm invisible})= 0.2000(6)$~\cite{pdg}, a significant fraction of such events will produce the monotop signature. Due to the larger partonic luminosity at the LHC, the $ug \to t Z$  process is expected to be more competitive with the corresponding decay channel (compared to $cg \to tZ$). We thus investigate the sensitivity of the monotop signature compared to existing experimental results using this mode.

Using existing experimental analyses we first derive nontrivial constraints on the parameter space of the models.  Single top searches at the Tevatron  and the LHC have not been optimized for the higher number of high $p_T$ jets compared to the SM. Nevertheless they may pose an important constraint.  On the other hand, the existing experimental search for monotops at the Tevatron~\cite{1202.5653} targets the hadronic top decay mode. While benefiting from  simpler hadronic signatures of one $b$-jet and two light jets as well as larger statistics, the information on the top charge is lost. We explore the benefits of employing the leptonic top decay signature 
at the LHC by suppressing otherwise dominant QCD multijet backgrounds and retaining information on the monotop charge production asymmetry, which is naturally expected to be large in some of the NP models under consideration due to the charge asymmetric $ug$ (versus $\bar u g$) initial state. 

As already mentioned, the production and detection of monotops have been studied before. Ref.~\cite{1106.6199} 
considered different scenarios, including one of the models studied in this work (an invisibly decaying or stable neutral vector boson). However, only the {\it hadronic} monotop signature was discussed,  this being the main difference with our work. In addition, the focus was on a much lighter invisible vector boson, with a mass of $50 ~\rm GeV$, leading to distinct kinematic features.  The discovery potential of hadronic monotop production was also discussed in Ref.~\cite{1308.3712}, where it was investigated within a model in which the top quark $A_{FB}$ arises from the on-shell production and decay of scalar top partners to top-antitop pairs with missing transverse energy~\cite{Isidori:2011dp}. On the other hand, the monotop signature with leptonically decaying top quarks was investigated in the context of R-parity-violating supersymmetry~\cite{Berger:1999zt,Berger:2000zk}. Although in these works the final state is the same as the one we are studying in this paper, there are two main distinctions: (1) the production scenarios studied are different; and (2) the previous analyses were targeting monotop production at the Tevatron. Consequently, the kinematical distributions  of the final state particles are distinct, as is the importance of the different backgrounds, leading in turn to a distinct problem and conclusions.  Finally, Ref.~\cite{1109.5963} studied both the monotop hadronic and leptonic signatures. However, we do not agree with this work on which are the main backgrounds, and as a consequence the conclusions obtained are different.

The paper is structured as follows: The example NP models are presented in Sec.~\ref{sec:2} and constraints on their parameter space from existing single- and monotop studies are derived in Sec.~\ref{sec:3}. In Sec.~\ref{sec:4} we present a new search strategy using the leptonic monotop decay signature and its discovery reach at the LHC. Finally, we summarize our conclusions in Sec.~\ref{sec:5}.

\section{The models}
\label{sec:2}

\subsection{$Z'$ model for asymmetric $t\bar t$ production}
We consider a model containing a $[\bar t u]$ flavored color- and weak-singlet $Z'$ vector boson, with a coupling to the right-handed up and top quarks~\cite{Jung:2009jz} (see also~\cite{Ko:2011di}). The relevant interaction Lagrangian is given by
\begin{equation}
\mathcal L^{}_{utZ'} = g_{utZ'}  \bar u \slashed {Z'} P_R t + {\rm h.c.} \,,
\label{eq:Lagr}
\end{equation}
where $P_{R} \equiv (1+\gamma_5)/2$. Note that we assume the $Z'$ is not self-conjugate in order to suppress same sign top production~\cite{Jung:2011zv}. Nonetheless, a self-conjugate $Z'$ would not modify the results of our analyses.

As discussed in the Introduction the main motivation for the model comes from its possible contributions to $t\bar t$ production at the Tevatron and the LHC. Namely, the exchange of the $Z'$ in the $t$-channel, due to its forward peaking,  leads to a positive $A_{FB}$ contribution, increasing with $m_{t\bar t}\equiv (p_t+p_{\bar t})^2$ and $|\Delta y_{t\bar t}| \equiv |y_t-y_{\bar t}|$ as observed by CDF and D\O.  It also produces a similarly positive contribution to $A_C$, in excess of the measurements. However, the associated production of the $Z'$ with a top-quark can produce an additional negative contribution to $A_C$.  Namely, the $Z' \to \bar t u$ decay yields a  $\bar t$  quark which tends to be boosted in the same direction as the incoming $u$ quark. Taking
into account the harder $u$ quark vs.~gluon parton distribution functions (PDF's) in the proton, one concludes that on average the $\bar t$ is produced with a larger rapidity than the $t$, thus yielding a negative contribution to $A_C$. At the LHC, the cross-section for the CP conjugate
process, $\bar u g \to Z'^\dagger \bar t \to \bar u t \bar t$, is typically an order of magnitude smaller, due to the $\bar u$-quark PDF in the initial state.  At the Tevatron, associated production of the vector mediators produces a negative contribution to $A_{FB}$.	However,	this effect is suppressed relative to the positive $A_{FB}$ contribution from $Z'$ $t$-channel exchange by the smaller gluon vs.~$u$-quark PDF's inside the proton at the lower collider energy.

The $Z'$ model is subject to a number of collider and low energy constraints.  In particular, atomic parity violation (APV) measurements are sensitive to one-loop induced $\bar u \slashed Z u$ vertex corrections~\cite{Gresham:2012wc}. However,  at the level of an effective theory in which one only considers the effects of interactions present in eq.~\eqref{eq:Lagr}, these constraints are rather weak (see~\cite{1209.4872} for a more detailed discussion of APV constraints also within possible UV completions of the effective model). On the other hand, measurements of $t\bar t j$ production and especially $tj$ resonance searches already put non-trivial constraints on the viable parameter space of the model (addressing the $A_{FB}/A_C$ discrepancy).  In particular, combined with $t\bar t$ observables they single out a range of $Z'$ masses $m_{Z'} \sim (200-300)$~GeV and $Z' \to \bar t u $ branching fractions $\mathcal B (Z' \to \bar t u ) \sim (1/3 - 1/4)$\,~\cite{1209.4354,1209.4872}.

The favored range for $\mathcal B(Z' \to  \bar t u)$ raises an immediate question: what are the viable candidates for the missing dominant $Z'$ decay? Possibilities include invisible decays, decays to quark or lepton pairs, and even more complicated decay chains possibly involving new intermediate particles. 
The second and third options result in a $t + n$ prong final state ($n \geq 2$). Here we focus on the first option which yields monotop events.

\subsection{$\Delta T=1$ weak FCNCs}
Instead of introducing a new massive neutral vector boson mediating top FCNCs ($Z'$ in previous section), one can also imagine the SM $Z$ boson acquiring flavor violating couplings to the top. Such weak FCNCs in the up-quark sector are highly suppressed in the SM, but are expected to be enhanced in many models of NP. Contrary to transitions among the first two quark generations, $\Delta T=1$ processes cannot be probed directly by low energy precision flavor experiments. If the associated NP scale is above the EW scale, currently being probed directly at the LHC, the experimental constraints are independent of the NP model details and the new effects can be efficiently parametrized in terms of a few lowest dimensional effective operators ($\mathcal Q_i$) involving only SM fields 
\begin{equation}
\mathcal L = \mathcal L_{SM} + \sum_i \frac{C_i}{\Lambda^{(d-4)}} \mathcal Q_i\,,
\end{equation}
where $d \equiv {\rm dim}(\mathcal Q_i)$ is the operator dimension.
In such an effective field theory (EFT) approach, the weak-scale dynamics should be described in a $SU(2)_L$ invariant way~\cite{Buchmuller:1985jz} leading to important correlations and constraints in particular from B physics on what top FCNCs are allowed~\cite{Fox:2007in}.  For example, among the lowest dimension ($d=6$) operators mediating $\Delta T=1$ weak FCNCs, only three remain virtually unconstrained by precision flavor data and can still be expected to yield significant contributions to FCNC top decays at the LHC,
\begin{align}
\mathcal Q^{w,i}_{LR} & = g \bar Q_3 \sigma^{\mu\nu} \tau^a \tilde H W^a_{\mu\nu} u^i_R \,, & \mathcal Q^{b,i}_{LR} & = g' \bar Q_3 \sigma^{\mu\nu} \tilde H B_{\mu\nu} u^i_R \,, & \mathcal Q^{u,i}_{RR} & = i \bar t_R \gamma^{\mu} u^i_R H^{\dagger}\overleftrightarrow{D}\hspace{0in}^{\mu}H\,,
\end{align}
where $i=1,2$, $\tilde H \equiv 2 i \tau^2 H^*$, $H^{\dagger}\overleftrightarrow{D}\hspace{0in}^{\mu}H\equiv H^{\dagger
}\overleftarrow{D}\hspace{0in}^{\mu}H-H^{\dagger}\overrightarrow{D}%
\hspace{0in}^{\mu}H$ while in the up-quark mass eigenbasis $Q_3 \equiv (t_L,V_{CKM}^{tj} d^j_L)$, $u^1_R\equiv u_R$ and $u^2_R \equiv c_R$\,. After EWSB both $\mathcal Q^{b,i}_{LR}$ and $\mathcal Q^{w,i}_{LR}$ lead to comparable $t \to u^i Z$ and $t \to u^i \gamma$ rates, and better sensitivity at the LHC is expected to come from the later  processes~\cite{AguilarSaavedra:2004wm}.  On the other hand, $\mathcal Q_{RR}^{u,i}$ only contributes to $t \to u^i Z$. Its contributions to trilinear vertices can be described by the effective Lagrangian
\begin{align}
\mathcal L^{u,i}_{RR} &= g_{tZu^i} \bar t \slashed Z P_R u^i + {\rm h.c. } + \ldots\,,
\label{eq:lutzi}
\end{align}
where $g_{tZu^i} \equiv C^{u,i}_{RR} \,g \,v_{\rm EW}^2/\cos \theta_W \Lambda^2$ while the dots denote additional terms involving the physical Higgs boson. At the LHC, the same interactions also lead to associated $tZ$ production through $g u^i$ scattering. Then, the substantial invisible decay width of the $Z$ produces the monotop signature. In the present paper we thus explore the sensitivity of the single- and monotop searches in constraining $\Delta T=1$ weak FCNCs mediated by the effective interaction in Eq.~\eqref{eq:lutzi}\,.

\subsection{Type III THDM with scalar DM}\label{sec:thdmiii}
Our final example involves scalar mediated top FCNCs and is based on a type III Two-Higgs-Doublet-Model (2HDM-III) supplemented by an extra singlet scalar field. The detailed structure of the model can be found in~\cite{Li:2011ja}. The particle content consists of SM fermions, two Higgs doublets, $H_1$ and $H_2$, and a real scalar $S$. 
The singlet $S$ is assumed to be $Z_2$ odd and is identified as a DM candidate. The Yukawa interactions of the two doublets are assumed to be of generic III 2HDM type.  Without loss of generality, one can choose a basis where only one of the Higgs doublets ($H_1$) obtains a vacuum expectation value $v_{\rm EW} \simeq 174$~GeV.  In the vanishing $H_1 - H_2$ mixing limit we can identify the $125$~GeV scalar discovered at the LHC~\cite{Aad:2012tfa} with the neutral CP even component of $H_1$, its coupling to EW gauge bosons and fermions being SM Higgs-like. Thus, after EW symmetry breaking, the FCNC SM-DM interactions are mediated mostly by the second, heavier CP even scalar state in the model ($h_2$) and can be described by the following effective Lagrangian~\cite{1107.0623},
\begin{equation}
{\cal L}_{h_2}^{\tilde y}=  \sum_{ij} \left( \tilde y_u^{ij} \bar u^i P_R u^j h_2  + \tilde y_d^{ij} \bar d^i P_R d^j h_2  \right) + {\rm h.c.} + \lambda v_{\rm EW} h_2 S S,
\end{equation}
where the last term arises from $H_1^\dagger H_2 S^2 $.  In the vanishing $H_1-H_2$ mixing limit $h_2$ does not couple to $ZZ$ nor $W^+W^-$ pairs. Depending on the $h_2$ and $S$ masses and relative sizes of $\tilde y$ and $\lambda$, the $h_2$ decay width gets the largest contributions from decays to $SS$ or $q_i\bar q_j$ pairs. Since the effective $\tilde y^{ij}_{q}$ couplings in the quark mass eigenbasis arise after diagonalizing the quark mass matrices (and couplings to $h_1$), naturalness of the SM quark mass hierarchy would imply $|\tilde y^{ij}_q| \lesssim \sqrt{m_i m_j}/v_{\rm EW}$\,~\cite{Cheng:1987rs}. We note however that in principle larger values are also possible.  In fact, in explicit flavor models these bounds can be saturated for some of the couplings.  As an illustration we consider the structure of quark Yukawas due to spontaneously broken horizontal symmetries~\cite{FNM}. The quark fields carry	horizontal	charges $H(u^i_R), H(d^i_R), H(Q^i_L)$ (while $H_{1,2}$ and $S$ do not carry a horizontal charge) so that the $H_1$ Yukawas are given by
$y^{ij}_u \sim \lambda^{|H(Q^i_L)-H(u_R^j)|}\,,~y^{ij}_d \sim \lambda^{|H(Q^i_L)-H(d_R^j)|}$\,,
with the expansion parameter $\lambda \simeq \sin \theta_C = 0.23$ being the sine of the Cabibbo mixing angle. After EW symmetry breaking, the quark mass matrices are given by $m^{ij}_{d,u} = v_{\rm EW} y_{d,u}^{ij}$\,. An assignment of horizontal charges leading to phenomenologically satisfactory quark masses and the CKM matrix, is $H(\{Q^1_L, Q^2_L, Q^3_L; u^1_R, u^2_R, u^3_R; d^1_R, d^2_R, d^3_R\}) = \{3,2,0;-3,-1,0;-3,-2,-2\}$~\cite{Leurer:1992wg}\,. The horizontal symmetries then also fix the sizes of $\tilde y_{u,d}^{ij}$
\begin{equation}
\tilde y_u \sim \left( 
\begin{array}{ccc}
\lambda^6 & \lambda^4 & \lambda^3 \\
\lambda^5 & \lambda^3 & \lambda^2 \\
\lambda^3 & \lambda & 1 
\end{array}
\right)\,, ~
\tilde y_d \sim \left( 
\begin{array}{ccc}
\lambda^6 & \lambda^5 & \lambda^5 \\
\lambda^5 & \lambda^4 & \lambda^4 \\
\lambda^3 & \lambda^2 & \lambda^2 
\end{array}
\right)\,.
\label{eq:flavModel}
\end{equation}
In particular, the largest off-diagonal element is in the top-charm sector $|\tilde y_u^{tc}| \sim 0.2$\,.

For weak scale $h_2$ masses, the off-diagonal entries of $\tilde y_d^{ij}$ (and also $\tilde y_u^{uc}$, $\tilde y_u^{cu}$) are also severely constrained experimentally by the neutral meson oscillation measurements~\cite{Harnik:2012pb}. On the other hand, the indirect constraints on $\tilde y_u^{ut}$, $\tilde y_u^{tu}$, $\tilde y_u^{ct}$ and $\tilde y_u^{tc}$ from $D^0$ oscillations are weaker
\begin{align}
|\tilde y_u^{ut} \tilde y_u^{ct} |, |\tilde y_u^{tu} \tilde y_u^{tc}| & < 0.030 \times \left(  \frac{m_{h_2}}{250\rm GeV} \right)^2\,, \nonumber \\ |\tilde y_u^{tu} \tilde y_u^{ct} |, |\tilde y_u^{ut} \tilde y_u^{tc}| & < 0.0088 \times \left(  \frac{m_{h_2}}{250\rm GeV} \right)^2\,, \nonumber\\
\sqrt{|\tilde y_u^{ut} \tilde y_u^{tu} \tilde y_u^{ct} \tilde y_u^{tc}|} & < 0.0036 \times \left(  \frac{m_{h_2}}{250\rm GeV} \right)^2\,,
\end{align}
and not yet probing their natural values (e.g. in~\eqref{eq:flavModel}). In any case, given these estimates for $h_2$ masses above the $SS$ and below the $t\bar t$ thresholds ($2m_S < m_{h_2}\lesssim 2 m_t$), and for $\lambda = \mathcal O(1)$ (consistent with obtaining the correct relic DM abundance~\cite{Li:2011ja,Greljo:2013wja}), the $h_2$ width will be naturally saturated by $h_2 \to SS$ decays. For $m_{h_2} < m_t$, the FCNC top decay $t \to c(u) SS$ might give competitive constraints on the model~\cite{Li:2011ja}. However, this mode quickly becomes ineffective for heavier $h_2$. In the following we therefore study the existing and prospective future constraints on the model using associated $t h_2$ production at the LHC, { for masses $m_{h_2}\gtrsim 150$~GeV} and assuming $\mathcal B(h_2 \to SS)\simeq 1$.\footnote{Nonetheless, our results can easily be rescaled to any value of $\mathcal B(h_2 \to SS)$.}

\section{Constraints from existing analyses} \label{cdf&atlas}\label{sec:3}

In this section we investigate, for each model, the bounds imposed by existing experimental analyses.  We compare their effectiveness in constraining the models' parameter space. Finally, we define useful benchmarks for studying the reach of our proposed monotop search strategy. 

\subsection{$Z'$ Model}\label{sec:Zpexist}
\begin{figure}[!h]
\begin{center}
\includegraphics[width=0.6 \textwidth]{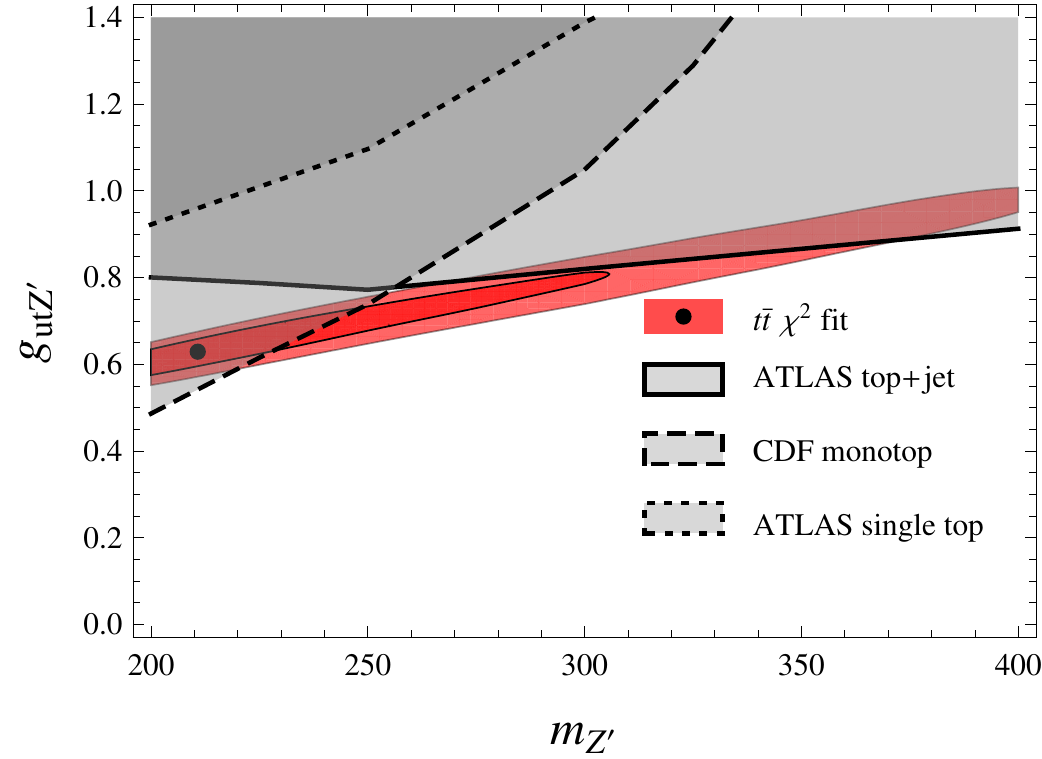}
\caption{\footnotesize [Color online] Constraints on the two dimensional $Z^\prime$ model parameter space with renormalization and factorization scales set to $\mu = m_t$ and $\mathcal{B}(Z'\to \mathrm{invisible})=3/4$. The black dot and the red regions represent the best fit point and the $1\sigma$ and $2\sigma$ preferred regions based on various $t\bar t$ observables (see text for details). Gray regions correspond to parameter space excluded at $95\%$ C.L. by ATLAS top+jet resonance search~\cite{atlastj} (full contour), CDF monotop search~\cite{1202.5653} (dashed contour) and ATLAS t-channel single top measurement~\cite{1205.3130} (dotted contour). }
\label{Fig:final_plot}
\end{center}
\end{figure}

Recently, a comprehensive analysis of the $Z'$ model in $t\bar t$ production has been performed in Refs.~\cite{1209.4354, 1209.4872}. We first update those results by including the latest experimental data and also an additional observable in the $\chi^2$ fit of $t\bar{t}$ phenomenology at the Tevatron and the LHC. In particular, the black dot and the red regions in Fig.~\ref{Fig:final_plot} respectively, show the best fit point and the $1\sigma$ and $2\sigma$ preferred regions of ($m_{Z'}, g_{utZ'}$) obtained through a $\chi^2$ fit of the following observables and their experimental values: 
inclusive Tevatron forward-backward asymmetry $A_{FB}$, for which the naive average of CDF~\cite{CDF:2012xba} and D\O~measurements~\cite{Abazov:2011rq} is used; unfolded differentiated $A_{FB}^{\mathrm{low,high}} = A_{FB}(m_{t\bar t}\lessgtr 450\,\mathrm{GeV})$ provided by CDF~\cite{CDF:2012xba}; inclusive $t \bar t$ cross-section at Tevatron~\cite{CDF:ConfNote10926}; inclusive $A_C$ and $t\bar t$ cross-section at the LHC where a rough average of ATLAS~\cite{ATLAS:2012an,ATLAS:Public} and CMS~\cite{CMS:2012eua,CMS:top-11-024} measurements is used; and finally the differential cross-section in the highest $m_{t \bar t}$ bin reported by ATLAS~\cite{Aad:2012hg}. We note that this last observable represents a stringent test of $t-$channel models addressing the $A_{FB}$ puzzle, and on its own disfavours the $Z'$ model over the SM.  However, the discrepancy in $A_{FB}$ with the SM is more significant still. Therefore, a fit including both of these observables (which seem somewhat incompatible from both $Z'$ model or SM point of view) favours the $Z'$ explanation over the SM in the regions marked by the ellipses in Fig.~\ref{Fig:final_plot}. In addition, the black curve in the plot shows the region compatible with the ATLAS bound on $tj/ \bar t j$ resonance production~\cite{atlastj}. Here and throughout this work we use $\mathcal B(Z'\to \mathrm{invisible})=3/4$ for definiteness, while the factorization and renormalization scales are set to $\mu=m_t$~\footnote{For the details on all our numerical calculations we refer the reader to Sec.~\ref{event_selection}.}. In our theoretical predictions to be compared with the experimental measurements, LO NP contributions are combined with the latest available (N)NLO SM predictions in the same manner as described in Ref.~\cite{1209.4872}.

We extend the above $Z'$ model analysis by including a direct and important consequence of $\mathcal B(Z'\to \mathrm{invisible})>0$: the prediction of monotop production. In the following we compare the bounds coming from the {\it t-}channel single top production measurement at ATLAS~\cite{1205.3130}, and the limits given by the recent monotop search of CDF~\cite{1202.5653}. 

Single top quarks are produced via three different processes in the SM: a {\it t-}channel $W$ boson exchange inducing the quark-level transition $q b \to q' t$~\cite{1103.2792}, dominant at both the LHC and the Tevatron; a $W t$ associated production via $bg$ fusion~\cite{1005.4451}; and $t\bar b$ production via $W$ exchange in the {\it s-}channel~\cite{1001.5034}. The single top production signature in the $Z'$ model is different from all the three SM processes because it is given by a single top quark together with additional missing energy but no extra charged tracks nor (light or $b$-) jets. Consequently one can expect that existing measurements targeting SM single top production will not be optimized for the $Z'$ mediated process. We quantitatively investigate this issue by deriving the constraint on the model coming from the ATLAS collaboration measurement of the {\it t-}channel single top production cross-section using $1.04 ~\rm fb^{-1}$ of $pp$ collision data at $\sqrt{s}=7 ~ \rm TeV$ \cite{1205.3130}.

In order to estimate the bounds coming from this analysis, we have simulated, using {\sc MadGraph5 \& MadEvent}~\cite{MG} (MGME) with the same simulation parameters as in~\cite{1209.4872}, the signal within the $Z'$ model, i.e, $pp \to {t}Z^{\prime}, Z'^{} \to \slashed E_{T} $, for masses of the $Z'$ and $g_{utZ'}$ in the same ranges as those in Fig.~\ref {Fig:final_plot}. Repeating the event selection in Ref.~\cite{1205.3130} for our signal events and performing a $\chi^{2}$ test with the four observables that appear in Table 1 in that work (requiring the p-value to be greater than 0.05), we have selected the points in parameter space which are in agreement with ATLAS results at the $95 \%$ confidence level (C.L.). The exclusion limits based on this analysis in the $m_{Z'}$ vs. $g_{utZ'}$ plane in parameter space are shown in Fig.~\ref{Fig:final_plot} (dotted contour) . 

More competitive contraints can be derived using the monotop search. Recently, the CDF experiment performed the first search for monotops through the production of a dark matter candidate (D) in association with a top quark, using $7.7 ~\rm fb^{-1}$ of $p\bar p$ collision data at $\sqrt{s}=1.96 ~ \rm TeV$ \cite{1202.5653}. The analysis considers exclusively the hadronic decay mode of the top quark, yielding a final state of three jets with missing transverse energy. The observed data was found to be consistent with SM backgrounds' expectations, and 95\% C.L. upper limits were set on the cross-section of $p \bar p \to D+t$ in the D mass range $0-150 ~\rm GeV$.

In order to estimate the limits coming from this CDF monotop search, we proceed, as described before, by simulating the signal within the $Z'$ model for $\sqrt{s}=1.96 ~ \rm TeV$. We keep the events passing the CDF experimental cuts and, using the maximum likelyhood method~\cite{pdg}, select  points in parameter space which are consistent with CDF results at the 95\% C.L.. The exclusion limits based on this search are also shown in Fig.~\ref{Fig:final_plot} (dashed contour).

Comparing the exclusion regions based on both analyses we note that the ATLAS single top analysis does not constrain the $2\sigma$ $t\bar t$ preferred parameter region in Fig.~\ref{Fig:final_plot} while the CDF monotop search sets a significant bound on the model, ruling out a part of the otherwise preferred parameter space. This is mainly due to the fact that the signal studied by the CDF analysis matches closely the one we are investigating within the $Z'$ model, contrary to the ATLAS case focusing on SM single top production. Thus the CDF search reach is much larger despite the fact that the ATLAS single top analysis is based on $pp$ collisions at considerably higher energy yielding larger single- and monotop event samples.

\subsection{$Z$ mediated $\Delta T=1$ FCNCs}\label{sec:ZFCNCs}
Searches for $Z$-mediated FCNCs in top-quark decays have been performed both at the Tevatron and recently at the LHC. The latest search result targeting the $t \to Zq$ decays with a topology compatible with the decay chain $ t \bar t \to Wb+Zq \to \ell\nu b+\ell\ell q$ is due to CMS. At $\sqrt{s}=$~8~\rm TeV and using an integrated luminosity of 19.5 $\rm fb^{-1}$ they find $\mathcal B(t\to Zq)< 0.07 \%$ at 95 \% C.L.~\cite{cmsFCNC}.

This bound on the branching ratio can be translated into constraints for the $tZu$ and $tZc$ couplings in Eq.~\eqref{eq:lutzi}. For simplicity we will assume $\mathcal{B}(t \to Wb) + \sum_{u^i=u,c} \mathcal{B}(t \to Z u^i) =1$ (i.e. no other new decay channels of the top are significant) and $|V_{CKM}^{tb}|\simeq 1$ as strongly indicated by the global CKM fits~\cite{pdg}. Defining
\begin{equation}
\rho_{WZ} \equiv \frac{(2 m_W^2 + m_t^2) \left(1-\frac{m_W^2}{m_t^2}\right)^2}{(2 m_Z^2 + m_t^2) \left(1-\frac{m_Z^2}{m_t^2}\right)^2} \simeq 1.08,
\end{equation}
which takes into account the dominant phase-space difference in $t\to Wb$ and $t\to Zq$ decays neglecting $b$ and lighter quark masses, we can write
\begin{equation}
{\mathcal{B}}(t \to Zq) = \left({1+ \frac{m_Z^2}{ v_{EW}^2} \frac{\rho_{WZ}  }{ \sum_{u^i=u,c} | g_{tZu^i}|^2}}\right)^{-1}\,.
\end{equation}
Numerically, the CMS search limit on the branching ratio results in constraints for the $tZu^i$ couplings given by ${g_{tZu^i}} < 0.014$ at 95 \% C.L.\,. 

Next, we should compare this bound with the results from existing single- and monotop searches as in the previous subsection. Incidently the FCNC $tZu$ interaction in Eq.~\eqref{eq:lutzi} coincides with the $Z'$  model interactions in Eq.~\eqref{eq:Lagr}. Thus, we can employ the CDF monotop search results from the previous subsection directly by choosing the appropriate $Z'$ mass ($m_{Z'} = m_Z$)\,. In this way we obtain a bound of ${g_{tZu}} < 0.3$ at 95 \% C.L.\, (the ATLAS single top measurement again yields an even weaker constraint). 
Furthermore, the corresponding  bounds on $g_{tZc}$ are worse still due to the suppressed charm PDF in the proton.
Thus, we observe that in the case of $Z$ mediated $\Delta T=1$ FCNCs, the existing monotop search at the Tevatron is not competitive with the latest LHC analyses employing the $t\to Zq$ decay. 

\subsection{2HDM-III + DM}
\begin{figure}[!h]
\begin{center}
\includegraphics[width=0.6 \textwidth]{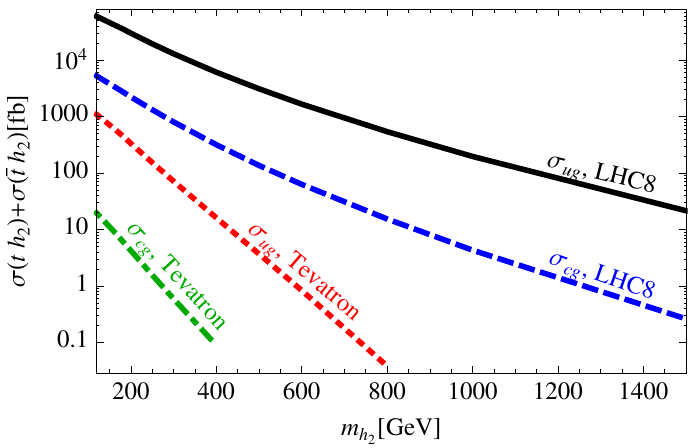}
\caption{\footnotesize [Color online] Partonic monotop production cross-sections in the THDMIII+DM at the Tevatron and the LHC, normalised to the invisible branching fraction of $h_2$ and its relevant couplings $\tilde y^{ij}_u$, as a function of $h_2$ mass (see text for details). For the Tevatron, the $cg$ and $ug$ fusion induced contributions are shown in dot-dashed (green) and dotted (red) lines, respectively. For the 8~TeV LHC, these same contributions are shown in dashed (blue) and full (black) lines, respectively.}
\label{fig:TopHiggsFCNC}
\end{center}
\end{figure}

The monotop production cross-section in the 2HDM-III + DM can be mediated by any of the couplings $\tilde y^{tc}_u,\tilde y^{ct}_u$ (through the partonic process $cg \to t (h_2 \to SS)$ and its charged conjugate) and $\tilde y^{tu}_u,\tilde y^{ut}_u$ (through $ug \to t (h_2 \to SS)$ and its charged conjugate). Compactly it can be written as
\begin{equation}
\sigma_{\rm monotop } \simeq \sigma(t+h_2) + \sigma(\bar t + h_2) \simeq (|\tilde y_u^{tc}|^2 + |\tilde y_u^{ct}|^2) \sigma_{cg} + (|\tilde y_u^{tu}|^2 + |\tilde y_u^{ut}|^2) \sigma_{ug}\,,
\end{equation}
where in the first equality we have assumed $\mathcal B(h_2 \to SS) \simeq 1$. The expected hierarchy among the values of $\tilde y_u^{ij}$ means that the PDF suppressed $cg$ fusion proccess could easily dominate over the partly valence $ug$ process. To gauge the sensitivity of the existing single and monotop searches to these interactions, we plot in Fig.~\ref{fig:TopHiggsFCNC} the total normalized monotop production cross-sections ($\sigma_{cg}$ and $\sigma_{ug}$) at the Tevatron and the 8~TeV LHC (computed using MGME, with CTEQ6L~\cite{Pumplin:2002vw} PDFs and factorization and renormalization scales fixed to the top mass) as a function of the $h_2$ mass. 
 We immediately observe that for natural values of $\tilde y^{tc}_u,\tilde y^{ct}_u$ ($\tilde y^{tu}_u,\tilde y^{ut}_u$) and $m_{h_2}\gtrsim 150$~GeV  as discussed in Sec.~\ref{sec:thdmiii}, the expected number of monotop events in the complete Tevatron run II is below one. Thus, the existing monotop search by CDF cannot probe the interesting region of parameter space of the model.  On the other hand, the relevant cross-sections at the 8~TeV LHC are more than two orders of magnitude larger, with the partonic $cg$ fusion process becoming even more pronounced.  The existing LHC dataset could thus already exhibit potentially significant sensitivity to $\tilde y^{tc}_u$ (and $\tilde y^{ct}_u$).  We study this possibility in detail in the next section.

\section{Search strategy for leptonic monotops and its discovery/exclusion reach} \label{search_strategy}
\label{sec:4}

\subsection{Signal features and main backgrounds} \label{signal_bckg}
The main signatures associated with monotop production in the three models under study in this work, can be classified according to the main top quark decay chains,
\be
pp \to t + X \to b W + \slashed E_{T} \to (bjj + \slashed E_{T} \quad {\rm or} \quad b\ell + \slashed E_{T})\,,
\ee
where $X$ can stand for a $Z'$ coupling to $u$ and $t$, a SM $Z$ boson coupling to $u$ and $t$ or a $h_2$ scalar coupling to $c$ and $t$; $j$ and $b$ denote light/c- and b-jets, respectively, $\ell$ a charged lepton, and $\slashed E_{T}$, missing transverse energy. 

In the following we focus on the signal with the top quark decaying leptonically. There are two reasons for studying this mode instead of the hadronic one: firstly, as mentioned in the Introduction, the hadronic mode has already been largely explored. Secondly, the leptonic mode backgrounds are cleaner so that they can be simulated and controlled reliably. In particular, one can forego dealing with QCD multijet backgrounds which have large theoretical uncertainties and in general require data-driven methods to control. 

\begin{figure}[!ht]
\begin{center}
\includegraphics[width=0.9 \textwidth]{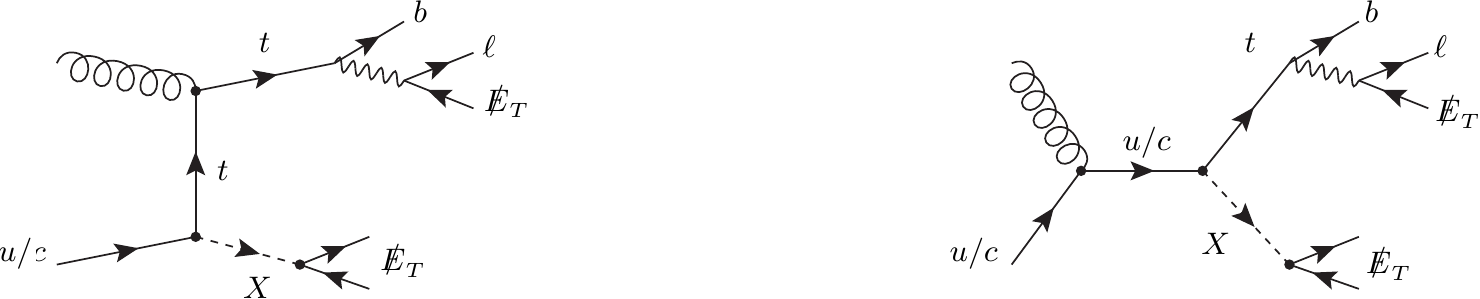}
\caption{\footnotesize { Leading order Feynman diagrams for the signal: $pp \to t + X \to b\ell + \slashed E_{T}$. $X$ represents a $Z'$ or a SM $Z$ boson both coupling to $u$ and $t$ or a $h_2$ scalar coupling to $c$ and $t$.  Note that the $\slashed E_{T}$ comes from the $X$ particle decay and the neutrino from the leptonic decay of the top quark.} }\label{feyn}
 \end{center}
\end{figure}

Since we are interested in the leptonic top decay mode, the topology of the sought signal for all models consists on one $b$-jet, a lepton, missing transverse energy associated to both the unobserved decay of the $X$ particle and the neutrino coming from the leptonic top decay, and light jets\,\footnote{Throughout this work we use the term {\it light jet} for all non-b-tagged jets.} from initial and final state radiation (ISR and FSR, respectively). Fig.~\ref{feyn} shows the leading order Feynman diagrams for the process $pp \to t+X \to \ell b+\slashed E_{T}$. { Note that due to the relevant PDFs, the LHC cross-sections associated with the conjugate diagrams are suppressed compared to those in Fig.~\ref{feyn} if the incoming parton is a $ u$ quark. This is not the case if the initial parton is a $ c$ quark because $c$ and $\bar c$ PDFs coincide.}

A distinctive characteristic of this signature is the excess of $\slashed E_{T}$ in the production of a single top. Nonetheless, the main discriminating variable of the leptonic monotop signature is related to the transverse mass of the lepton plus missing energy system (from now on we refer to it as $M_{T}$). This variable is defined as $M_{T}^{2}=(E_{T}(\ell)+\slashed E_{T})^{2}-(p_{x}(\ell)+\slashed E_{x})^{2}-(p_{y}(\ell)+\slashed E_{y})^{2}$. When $\ell$ and $\slashed E_{T}$ are the decay products of a particle with mass $M$, the $M_{T}$ spectrum has an end-point at $M_{T}^{max}=M$~\cite{pdg}. All the main backgrounds turn out to exhibit this feature. On the other hand, in the case of the signal, since there are two sources of missing energy, this is not the case. As we discuss in detail below, this is why $M_T$ turns out to be a key variable in distinguishing between signal and backgrounds. 

The dominant backgrounds (after cuts, as discussed below), in order of importance, arise from the SM processes of $t\bar t$, single-top, $W j$ (where $j$ can be a light or a heavy flavored jet), and diboson (VV)  production. In the following, we examine each of them separately emphasizing the role of $M_T$ in their reduction:

\begin{itemize}

\item{$t\bar t$:} The largest background comes from the SM production of $t\bar t$ pairs. Note that this is in contrast to leptonic monotops at the Tevatron studied in Refs.~\cite{Berger:1999zt,Berger:2000zk}, where the dominant background was $W j $, and is due to the fact that $t \bar t$ cross-section rises faster than $Wj$ when increasing collider energy.  In the semileptonic decay mode we expect the spectrum of $M_{T}$ to have an end point at $M_{T}^{max}=m_W$. However, this is not the case if there is missing energy coming from misreconstructed jets; $M_T$ becomes unconstrained. Moreover, in the dileptonic mode, if one of the leptons is missed, one is left with processes in which $M_{T}$ can again exceed $80 ~\rm GeV$. 

\item{\it single top}: One could expect single top to be the main background because it is irreducible, up to a jet that could come from ISR. However, the only single top process that can produce a large $M_{T}$ is the $tW$ associated production, i.e, $pp \to tW \to b \ell \ell \nu \nu$. If a lepton is missed, the $M_{T}$ is not constrained by $m_W$, and this process can contribute to the background. On the other hand, it has a low cross-section and thus turns out to be less important.

\item{\it Wj }: We study processes with up to three jets in the final state and include the production of $W$ in association with heavy flavored jets. These backgrounds are important mainly because of their large cross-sections, but turn out not to be the main background because of their small acceptances after cuts.

The largest contribution to the inclusive cross-section comes from the associated production of $W$'s with light jets. However, these processes need a fake $b$-tagged jet in order to contribute to the background. Since the $b-$mistag rate in current ATLAS and CMS analyses is of the order of $1/100$ and $1/1000$, depending on the working point of the $b$-tagging algorithm, this contribution to the background can be brought under control. Furthermore, the production of $W$ plus heavy flavored jets does not make an important part of the background either: $W b$ production, although being irreducible, has a small cross-section; while $W c$ requires a $c-b$-mistag, which is usually of the order of $\sim 1/20$, suppressing this background sufficiently. 

\item{$VV$}: This background is suppressed compared to the ones described above because of the difficulty of faking a $b\ell \slashed{E}_T$ final state. The process with the largest cross-section, $W^+ W^-$, can only contribute if one $W$ decays leptonically and the other one hadronically. In addition, a $b$-jet can only come from a mistag in one of the $W$ decays. Similarly, $W Z$ production also needs to involve a leptonic $W$ decay and either a mistagged jet (missing the other one) or a missing $b$ jet from the $Z \to b \bar b$ decay in order to contribute. Finally, the process with the smallest cross-section, $ZZ$, should have one of the bosons decaying leptonically and the other one hadronically in order to pass selection cuts. In this case, one of the leptons should be missed and again, there should be either a mistagged jet or a missing $b$ jet from the $Z$ hadronic mode. Apart from all this, values of $M_T>m_W$ are not likely to be produced and as a result this background becomes almost negligible.  

\item{\it QCD multijet}: We can neglect, as already argued, the background coming from QCD multijet production because in these processes reconstructed leptons can only come from misidentified jets.
In addition, large missing transverse energy can only come from high $p_{T}$ misreconstructed jets. Given the signal features, a high $p_{T}$ jet veto which suppresses the QCD missing energy coming from such misreconstructed jets turns out to be very effective in suppressing this background to the leptonic monotop signature. The details of such a jet veto are discussed below.

\item{} Finally, we also neglect the background coming from SM $Zt/Z \bar t$ production. Even though being a monotop signature itself when $Z \to \nu \bar \nu$, the inclusive SM cross-section is low ($\sim 0.24~\rm pb$ \cite{1302.3856}) and, after imposing the selection cuts discussed below, becomes negligible when compared to the rest of the backgrounds. 
\end{itemize}

In the next subsection we study the main features of the signal and the main backgrounds through Monte Carlo simulations and point out variables, useful in discriminating them from each other. This allows us to perform an event selection which optimizes, as discussed in Section~\ref{discovery_reach}, the discovery/exclusion reach of each model. 

\subsection{Event Generation and Selection}\label{event_selection}
The signal and backgrounds under study are modeled using MGME and {\sc Pythia}~\cite{pythia1,pythia2} for initial and final state radiation, parton showering and hadronization, as well as {\sc PGS} \cite{pgs} for detector simulation. We simulate collisions produced at the LHC, for an integrated luminosity of $21.7 \,\rm fb^{-1}$ and running at a center of mass energy of $\sqrt{s}=8 ~\rm TeV$, from now on referred to as the 2012 data. Since in all cases the simulated production processes are inclusive, we implement the MLM matching scheme in order to avoid double counting. All simulations employ the CTEQ6L~\cite{Pumplin:2002vw} PDFs with the renormalization and factorization scales  fixed to the top mass. In order to get accurate estimates of all the backgrounds we simulate ten times the expected 2012 data for each, except for $W j$ for which we simulate only twice the actual data.

All simulated background inclusive cross-sections are normalized to the most precise currently  known estimates. In particular,  the $t\bar t$ cross-section is normalized to the inclusive NNLO theoretical prediction~\cite{1303.6254}; for the single-top case, the cross-section is the sum of the NLO predictions of the $t$-, $s$- and $tW$ channels~\cite{1210.7813} with $W \to j j$ for the last process, while the $tW, W\to l \nu$ production is computed at LO. The $Wj$ cross-section is normalized to its experimentally measured value~\cite{cms_wj}, while $VV$ cross-sections correspond the theoretical NLO predictions~\cite{1105.0020}. 

We pre-select events by requiring exactly one {\it b}-tagged jet with $p_{T}> 25 ~\rm GeV$ and $|\eta|<2.5$ and one lepton (electron or muon) with $p_{T}> 20 ~\rm GeV$ and $|\eta|<2.5$. { Leptons must be isolated from jets by a cone of radius $\Delta R=0.4$ or else they are considered missed and jets are reconstructed using the anti-kt algorithm with a radius parameter of 0.4.}
With this selection, we study the spectrum of  $\slashed E_{T}$, $M_T$, and the $b$-jet and light jet multiplicities.  For the sake of simplicity, we show below results where the signal corresponds to that of the $Z'$ model (we refer to it as $X$ throughout this subsection), given that the distributions in all the three models are similar. 
  
\begin{figure}[!htbp]
\begin{center}
\begin{minipage}[b]{0.45\linewidth}
\begin{center}
\vspace{1cm}
\includegraphics[width=1 \textwidth]{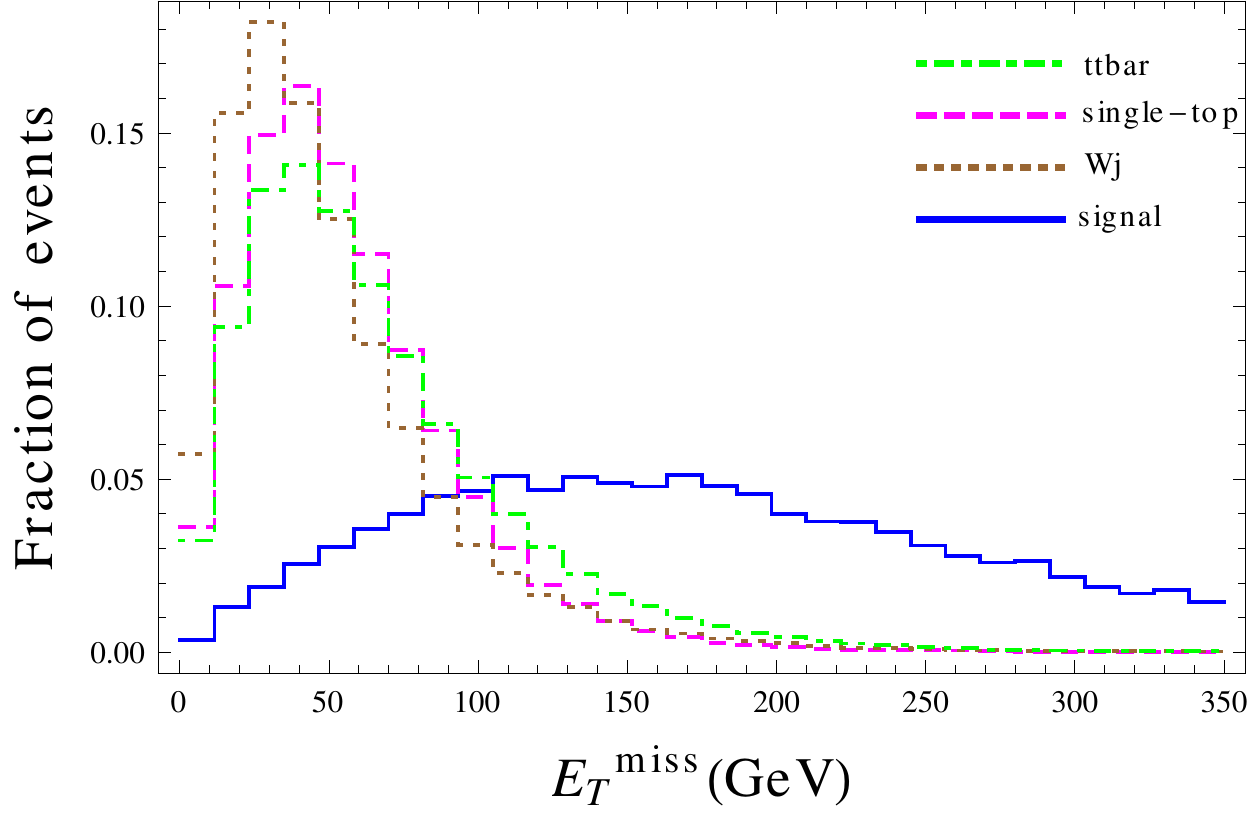}
\newline
(a)
\end{center}
\end{minipage}
\hspace{0.5cm}
\begin{minipage}[b]{0.45\linewidth}
\begin{center}
\includegraphics[width=1 \textwidth]{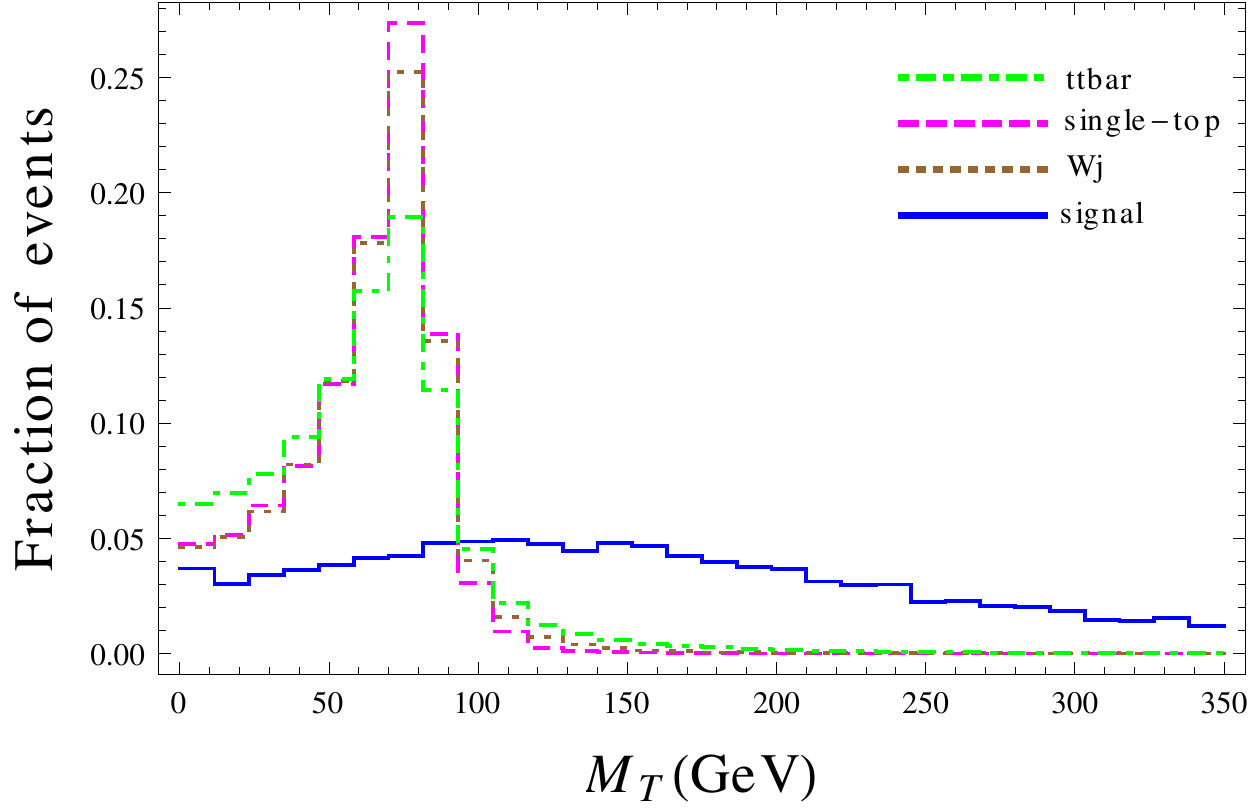}
\newline
(b)
\end{center}
\end{minipage}
\end{center}

\vspace{0.5cm}
\begin{center}
\begin{minipage}[b]{0.45\linewidth}
\begin{center}
\includegraphics[width=1 \textwidth]{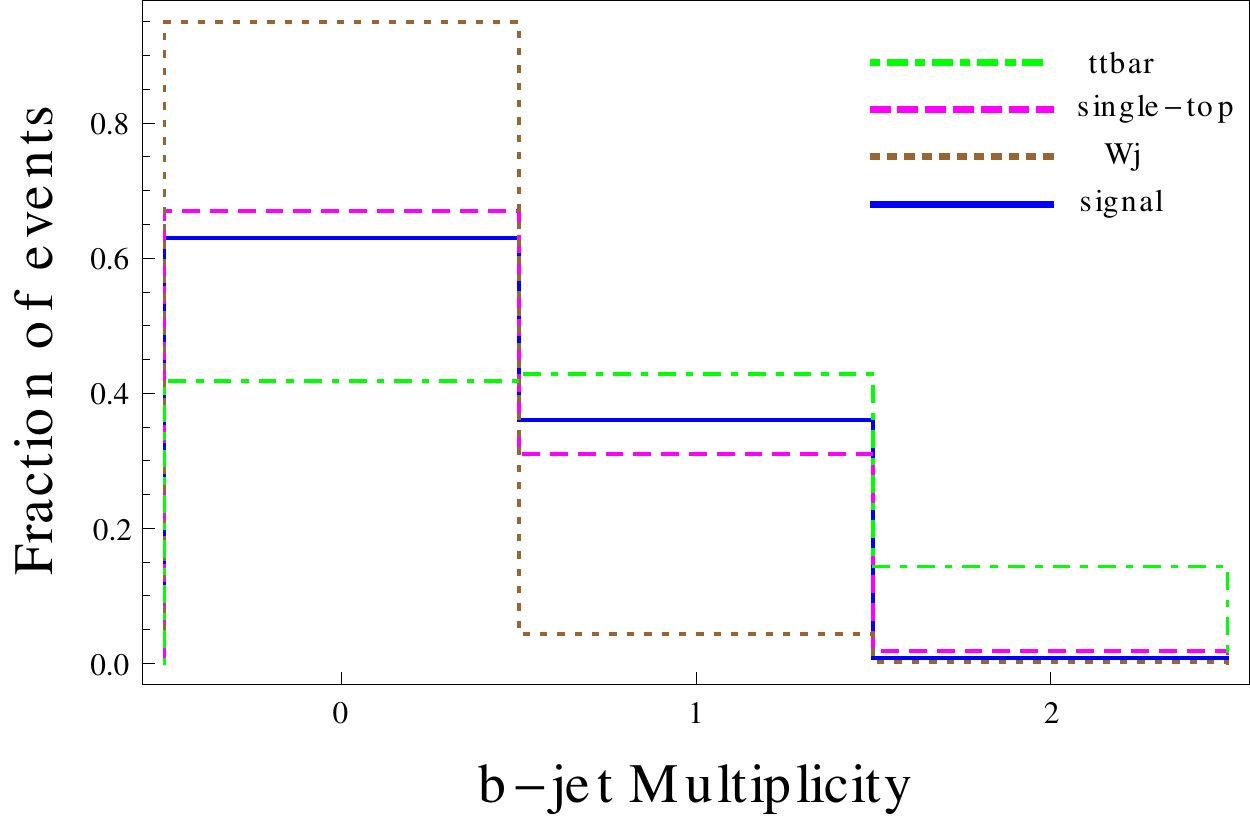}
\newline
(c)
\end{center}
\end{minipage}
\hspace{0.5cm}
\begin{minipage}[b]{0.45\linewidth}
\begin{center}
\includegraphics[width=1 \textwidth]{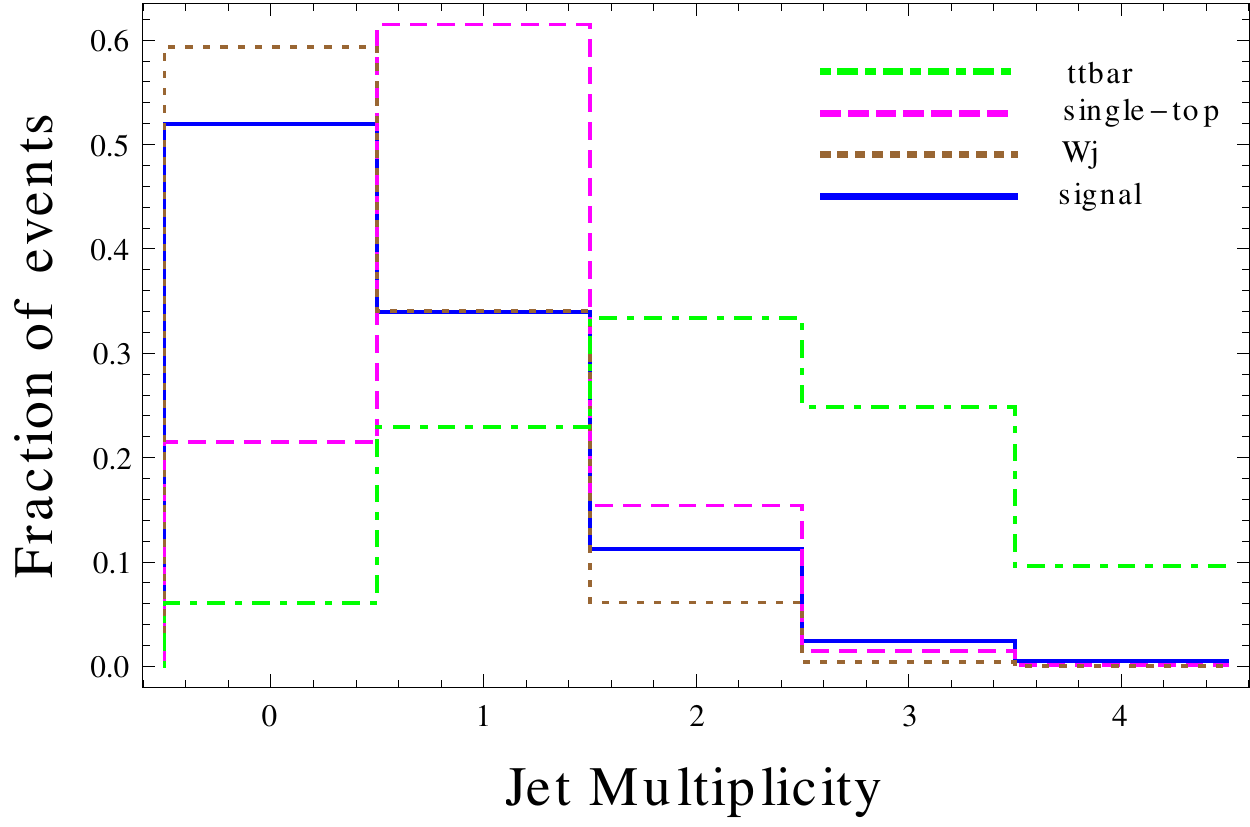}
\newline
(d)
\end{center}
\end{minipage}
\caption{\footnotesize{[Color online] Signal and background distributions of the following variables after the selection of events described in the text: (a) Missing transverse momentum, $\slashed E_{T}$, (b) Transverse mass of the lepton$+$missing energy system, $M_T$, (c) {\it b}-jet multiplicity,  (d) light jet multiplicity. The signal corresponds to leptonic monotop production ($t$ or  $\bar t$) at the LHC@8TeV in the $Z'$ model with $m_{Z'}=325 ~\rm GeV$. The signal spectrum of the other models is similar and is thus not shown.}}
\label{MET&MT}
\end{center}
\end{figure}

In Fig.~\ref{MET&MT} we present the spectra of the above mentioned variables for the main backgrounds ($t \bar t$, single-top and $W j$) and the signal: 
\begin{itemize}

\item Fig.~\ref{MET&MT} (a) shows the missing transverse energy spectrum where, as it can be seen, the background and signal can be discriminated clearly. Most of the background is concentrated in the region $\slashed E_{T} \lesssim (100-150) ~ \rm GeV$, while an important contribution from the signal is present also for larger values of this variable. 

\item Fig.~\ref{MET&MT} (b) shows the $M_{T}$ signal and the background distribution. As it was mentioned in Section~\ref{signal_bckg}, this is an interesting and useful variable to distinguish signal from background because once the pre-selection is made, all the backgrounds contain a $W$ and a unique source of missing energy, the neutrino, coming from its decay. The case of the signal is different because although only one $W$ is present, the spectrum is displaced to larger values of $M_{T}$  due to an additional missing transverse energy contribution coming from the $X$ particle production. As a result, and as seen in Fig.~\ref{MET&MT} (b),  a cut on $M_{T} \ge 80 ~ \rm GeV$ reduces the backgrounds considerably while conserving most of the signal.  It is worth noting at this point that cuts in $M_T$ end up having little correlation with cuts in $\slashed E_{T}$, as it will be shown in the cut-flow Tables below.

\item Figure~\ref{MET&MT} (c) shows the $b$-jet multiplicity. For this case, we only require events to have exactly one lepton. We can see in the figure that, as expected, selecting events with only one $b$-jet diminishes considerably ($\sim 10^{-2}$) the $W j$ background.

\item Finally, the signal and background light jet multiplicities are shown in Fig.~\ref{MET&MT} (d). As it can be seen, one can get rid of a good fraction of the $t\bar t$ background by imposing a veto on events with 2 or more light jets. As a matter of fact, one expects most semi-leptonic $t\bar t$ events to contain 2 light- and 2 $b$-jets. 
\end{itemize}

Having performed a characterization of the signal and backgrounds, in the next subsection we proceed to describe the cuts to be imposed on them in order to optimize the significance of the discovery/exclusion for each model. 

\subsection{Discovery/Exclusion reach} \label{discovery_reach}
Following the event selection analysis of the previous subsection 
we supplement our pre-selection of events (one lepton (electron or muon) with $p_{T}> 20 ~\rm GeV$ and $|\eta|<2.5$, one $b$-tagged jet with $p_{T}> 25 ~\rm GeV$ and $|\eta|<2.5$) with the additional  requirement of at most one jet with $25~{\rm GeV} < p_{T} < 120 ~\rm GeV$ and $|\eta|<4.5$. In the case of the $Z'$ model and the $utZ$ vertex, we also keep only the events with a {\it positive} lepton. In this way we get rid of half of the main background, ($t \bar t$), given that it is charge symmetric. On the contrary, most of the signal is kept since events with a negative lepton in the final state come from $\bar u g$ induced processes with PDF suppressed cross-sections. The same cut is not efficient for studying the $cg (\bar c g)$ induced processes which dominate in the THDMIII+DM, since these are charge symmetric at the LHC. Therefore in this case we keep  events with both positive and negative leptons. Finally, since QCD multijet background can only produce large missing transverse energy from high $p_{T}$ misreconstructed jets, we can suppress this background by controlling the number and energy of the jets.  For this reason, we also include an additional cut of $p_{T} < 120~ \rm GeV$ for extra light jets. 

After the pre-selection, we perform a cut-scanning on the following variables: $\slashed E_{T}$, $M_T$, the reconstructed top mass ($M_{b\ell\not E}$) \cite{topmass}, and two jet substructure variables: the number of tracks in the $b$-jet, and the $b$-jet mass. For the three models, we have found that the jet substructure variables are of little use to discriminate signal from background. We expect them to be more convenient in cases where the tops are boosted so that they can be mistagged as fat $b$-jets. This is not so in our case because we explore a signal region ($m_X \lesssim 400 ~ \rm GeV$) in which the tops are typically not too energetic. On the other hand, we have found that the reconstructed top mass hardly contributes to increase the signal significance due to the fact that there is a high correlation between this variable and $M_T$, and the latter being more sensitive.
 
We analyze the different cuts that come out of the cut-scanning by maximizing signal significance for each particular model's benchmark point.

\subsubsection{$Z'$ model}\label{Z'}
We propose a search strategy for the $Z'$ boson and investigate the model discovery/exclusion reach for the particular case of $\mathcal B(Z'\to \mathrm{invisible})=3/4$, $m_{Z'}=325 ~\rm GeV$ and $g_{utZ'}=0.7$, which is a representative point in the parameter space preferred by present data (see Fig.~\ref{Fig:final_plot}).

After performing the cut-scanning, the final event selection is given by: one positive lepton (electron or muon) with $p_{T}> 20 ~\rm GeV$ and $|\eta|<2.5$, exactly one {\it b}-tagged jet with $p_{T}> 25 ~\rm GeV$ and $|\eta|<2.5$, $\slashed E_{T} > 250 ~ \rm GeV$, $M_{T} > 120 ~ \rm GeV$ and up to one jet with $25~{\rm GeV} < p_{T} < 120 ~\rm GeV$ and $|\eta|<4.5$.

In Table~\ref{Table1} we present the signal and background cross-sections before and after these cuts are imposed. The last column shows the discovery/exclusion reach significance when only statistic uncertainties are taken into account.
\begin{table}[h!t]
\renewcommand{\arraystretch}{1.5}
{\tiny
\begin{center}
\begin{tabular}{|c|c|c|c|c||c|c|c|c|c||c|c|}
\hline
{\it $\ell^{+}$} & {\it b-}jets &  $M_T$ & $\slashed E_{T}$ & jets & $\sigma_{t\bar t j}$ & $\sigma_{tj}$& $\sigma_{Wj}$  &$\sigma_{VV}$ & $\sigma_{\rm signal}$& Sig. \\

($p_T > 20 ~\rm GeV$, & ($p_T > 25 ~\rm GeV$, & (GeV) & (GeV) & ($p_T < 120 ~\rm GeV$,& (pb) &(pb) &(pb)&(pb)& (pb) & ~   \\
$|\eta|<2.5$) &  $|\eta|<2.5$) & & & $|\eta|<4.5$) & & & & & &~ \\
\hline
- & - & - & - & - & 239 & 112.80 & 17035  & 84.20 & 43 &-\\ 
1 & 1  & $>$ 120  & $>$ 250 & $\leq$ 1 & $10^{-3}$ & 5.47*$10^{-5} $& $<$ 5.00*$10^{-5} $  &  1.43*$10^{-5}$  & 0.08 & 340 \\

\hline
\end{tabular}
\end{center}
}
\caption{\footnotesize{Signal and background cross-sections before and after the proposed cuts for the $Z'$ model are imposed. The signal is simulated for a reference point with $g_{utZ'}=0.7$, $m_{Z'}=325$ GeV and $\mathcal B(Z'\to \mathrm{invisible})=3/4$. The last column indicates the expected significance when only statistical uncertainties are taken into account.
}}
\label{Table1}
\end{table}
We observe that the signal for the chosen benchmark point is quite significant and easy to detect.  Moreover, we have verified that all the parameter space allowed by existing analyses (see Sec.~\ref{sec:Zpexist})  is accessible with this search. We conclude that the search strategy enhances considerably the visibility of the signal at the LHC compared to existing single top analyses. As the final significance suggests, a similar signal could be detectable even if suppressed by a factor of $\sim 100$. 

Motivated by this fact, we investigate the reach of the monotop leptonic search strategy in the  $g_{utZ'}-m_{Z'}$ plane and compare it with the ATLAS single-top analysis of Ref.~\cite{1205.3130}, but for the same luminosity and energy corresponding to the 2012 data.  We simulate the signal for masses in the range $[200,400] ~ \rm GeV$ and for a fixed coupling $g_{utZ'}$.  The number of signal events for different coupling values is then easily obtained since the signal cross-section scales as $g_{utZ'}^2 \times \mathcal B(Z'\to \mathrm{invisible})$. For the monotop search strategy, we keep the background and signal events passing the event selection described above and find, for each mass, the coupling for which the signal significance exceeds $2\sigma$.  In the case of the single top analysis, we simulate all the backgrounds considered in Ref.~\cite{1205.3130} for 8 TeV except QCD multijets (This means that the real single-top reach should be worse than what we actually find.). We keep the background and signal events passing the ATLAS event selection and find those points in parameter space where signal significance again exceeds $2\sigma$.  

\begin{figure}[!h]
\begin{center}
\includegraphics[width=0.5 \textwidth]{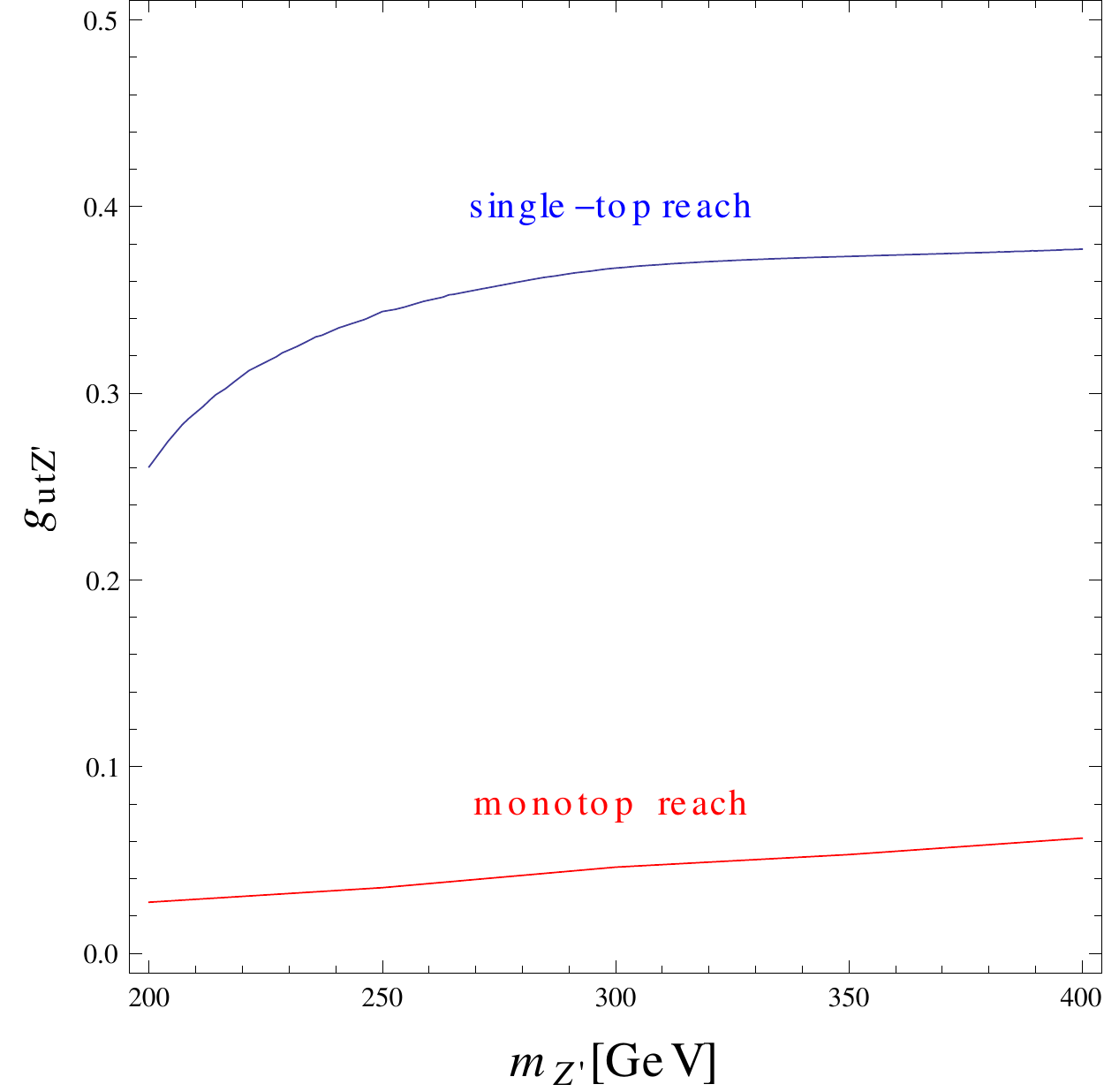}
\caption{\footnotesize  [Color online] Projected leptonic monotop and single top search strategies' reach with 2012 data in the  $g_{utZ'}-m_{Z'}$ parameter plane for the $Z'$ model. The single top search strategy corresponds to the one used in Ref.~\cite{1205.3130}.}
\label{monotop_single_reach}
\end{center}
\end{figure}

We present the results in Fig.~\ref{monotop_single_reach}, where both curves corresponding to each search strategy reach are shown. As it can be seen, the monotop reach is significantly larger than the single top one, making the presented monotop leptonic search strategy considerably more advantageous in the detection of the signal compared to the single top one. Note that for larger $m_{Z'}$ masses, the monotop leptonic search strategy improves the existing single top one by more than an order of magnitude in the coupling.  As a matter of fact, these results suggest that the single top search strategy is rather insensitive to the monotop signature. Nonetheless, notice that the $1\sigma$ and $2\sigma$ preferred regions in Fig.~\ref{Fig:final_plot} are above both curves, i.e, can be excluded by both searches.

\subsubsection {$Z$ mediated $\Delta T=1$ FCNCs}
Next, we consider the monotop leptonic search strategy for probing $Z$ mediated $\Delta T=1$ FCNCs. As discussed in Sec.~\ref{sec:ZFCNCs} the lightness of the $Z$ boson in comparison with the $t$ quark, makes FCNC top decays, $t\to Zq$, where $q=u,c$, a primary search channel. Currently the exclusion benchmark is set by the most stringent limit on $\mathcal B(t\to Zq) < 0.07\%$ at 95\% C.L.~\cite{cmsFCNC}, corresponding to an effective right-handed FCNC $u^i tZ$ coupling $g_{tZu^i}<0.014$ (see Sec.~\ref{sec:ZFCNCs}). As we show below, the monotop leptonic search strategy could potentially match this sensitivity (for the $u tZ$ coupling).  To this end we first study the leptonic monotop search reach considering only statistical uncertainties, but then also analyze the effects of systematic uncertainties. This additional step allows us to make a more accurate evaluation of the monotop search sensitivity to $Z$ mediated $\Delta T=1$ FCNCs compared to studies of FCNC top decays~\cite{cmsFCNC}.   

In this case the missing energy of the monotop signal comes from the $Z\to \nu\bar \nu$ decay together with the neutrino from the leptonic top decay. The cut-scanning for this scenario yields the largest significance for the following event selection: one positive lepton (electron or muon) with $p_{T}> 20 ~\rm GeV$ and $|\eta|<2.5$, exactly one {\it b}-tagged jet with $p_{T}> 25 ~\rm GeV$ and $|\eta|<2.5$, $\slashed E_{T} > 90 ~ \rm GeV$, $M_{T} > 110 ~ \rm GeV$ and at most one jet with $25 ~{\rm GeV} < p_{T} < 70 ~\rm GeV$ and $|\eta|<4.5$. By requiring the significance to reach $2 \sigma$, we find the lowest bound on $g_{utZ}<0.062$, corresponding to $\mathcal B(t\to Z u)<1.3\%$. Although this bound is weaker than the one obtained from FCNC top decays~\cite{cmsFCNC}, the monotop signature (especially if the leptonic and hadronic~\cite{1106.6199,1308.3712,1109.5963} signatures were combined) nonetheless appears to be interesting also for studying FCNC $utZ$ interactions. { We also note that repeating the procedure for the existing single top strategy~\cite{1205.3130} extrapolated to 2012 data also in this case yields a weaker limit of $g_{utZ}<0.14$}

We present in Table~\ref{Table2}, a cut-flow table containing the cross-sections of the signal (at fixed $g_{utZ}=0.062$) and main backgrounds obtained after imposing each of the cuts mentioned above. The last column shows the discovery/exclusion reach significances that result from each cut when only statistical uncertainties are taken into account. 
\begin{table}[!h]
\renewcommand{\arraystretch}{1.5}
{\tiny
\begin{center}
\begin{tabular}{|c|c|c|c|c||c|c|c|c|c||c|c|}
\hline
{\it $\ell^{+}$} & {\it b-}jets &  $M_T$ & $\slashed E_{T}$ & jets & $\sigma_{t\bar t j}$ & $\sigma_{tj}$ &  $\sigma_{Wj}$ & $\sigma_{VV}$ & $\sigma_{signal}$& Sig.\\

($p_T > 20 ~\rm GeV$, & ($p_T > 25 ~\rm GeV$, & (GeV) & (GeV) & ($p_T < 70 ~\rm GeV$,& (pb) &(pb) &(pb)&(pb)& (pb) &~ \\
$|\eta|<2.5$) &  $|\eta|<2.5$) & & & $|\eta|<4.5$) & & & & & &~ \\
\hline
- & - & - & - & - & 239 &  112.80 & 17035  & 84.20 & 0.90 &-\\ 

1 & - & - & - & - & 31.60 & 18.43  & 1827.27  & 8.86 & 0.10 & 0.34 \\

1 & 1 & - & - & - & 12.65  & 5.54 &  35.06 & 0.30 & 0.03 & 0.54 \\

1 & 1 & $>$ 110 &- & - & 0.87 &  0.10 & 0.18  & 3.60*$10^{-3}$ & 0.01 & 1.60  \\

1 & 1 & $>$ 110  & $>$ 90& - & 0.44 &  2.60*$10^{-2}$ & 7.44*$10^{-3}$ & 9.00*$10^{-4}$ & 9.00*$10^{-3}$ & 1.92  \\

1 & 1  & $>$ 110  & $>$ 90 & $\leq$ 1 & 0.15 & 1.60*$10^{-2}$ & 2.12*$10^{-3}$  & 7.50*$10^{-4}$  & 5.70*$10^{-3}$ & 2.05  \\

\hline
\end{tabular}
\end{center}
}
\caption{\footnotesize{Cut-flow table for the proposed cuts for the FCNC $utZ$ interaction (at fixed coupling $g_{utZ}=0.062$, see text for details). The last column indicates the expected significance when only statistical uncertainties are taken into account. }}
\label{Table2}
\end{table}
It is clear from Table~\ref{Table2} that $M_T$ is a key variable to suppress the backgrounds without loosing much of the signal, as it was largely discussed in Section~\ref{event_selection}, resulting in a sizeable enhancement in the signal significance. Also reflected in Table~\ref{Table2} is the fact that, as  expected, $\slashed E_{T}$ is a useful variable as well, given the excessive missing energy in the signal. Finally, as it was pointed out previously, it can be seen that there is little correlation between $M_T$ and $\slashed E_{T}$.

Next we discuss the effect of systematic uncertainties on the results of the analysis. Following~\cite{cmsFCNC}, we introduce a systematic uncertainty of 20\% for all background processes. 
After performing the cut scanning to optimize the cuts that result in the largest signal significance, we find a similar reach compared to the previous case. There is however a significant difference in the cuts imposed to the events: one positive lepton (electron or muon) with $p_{T}> 20 ~\rm GeV$ and $|\eta|<2.5$, exactly one {\it b}-tagged jet with $p_{T}> 25 ~\rm GeV$ and $|\eta|<2.5$, $\slashed E_{T} > 250 ~ \rm GeV$, $M_{T} > 110 ~ \rm GeV$ and up to one jet with $25~{\rm GeV } <  p_{T} < 120 ~\rm GeV$ and $|\eta|<4.5$. Most importantly, the cut on $\slashed E_{T}$ is  considerably strengthened while the jet $p_T$ cut is relaxed. 

As could be expected, in this case the lowest bound that could be obtained on $g_{utZ}< 0.077$ is slightly weaker than in the case when only statistical uncertainties are taken into account. In fact, the final number of signal and background events obtained after the cuts in Table~\ref{Table2} are applied, are 125 and 3930, respectively. When systematic and statistic uncertainties are both included, the preferred cuts leave only a few signal and background events: 11 and 18, respectively.  Results obtained when also systematic uncertainties are considered can be found in Table~\ref{Table3}. 

 \begin{table}[h!t]
\renewcommand{\arraystretch}{1.5}
{\tiny
\begin{center}
\begin{tabular}{|c|c|c|c|c||c|c|c|c|c||c|c|}
\hline
{\it $\ell^{+}$} & {\it b-}jets &  $M_T$ & $\slashed E_{T}$& jets & $\sigma_{t\bar t j}$ & $\sigma_{tj}$ & $\sigma_{Wj}$  &$\sigma_{VV}$ & $\sigma_{signal}$& Sig. \\

($p_T > 20 ~\rm GeV$, & ($p_T > 25 ~\rm GeV$, & (GeV) & (GeV) & ($p_T < 120 ~\rm GeV$,& (pb) &(pb) &(pb)&(pb)& (pb) &  \\
$|\eta|<2.5$) &  $|\eta|<2.5$) & & & $|\eta|<4.5$) & & & & & &~ \\
\hline
- & - & - & - & - & 239 &  112.80 & 17035  & 84.20 & 1.40 &-\\ 
1 & 1  & $>$ 110  & $>$ 250 & $\leq$ 1 & $10^{-3}$ & 5.5 *$10^{-5} $& $<$ 5*$10^{-5} $  & 1.4*$10^{-5}$  & 5.75*$10^{-4}$ & 2.01 \\

\hline
\end{tabular}
\end{center}
}
\caption{\footnotesize{Signal and background cross-sections before and after the proposed cuts for the FCNC $utZ$ interaction (at fixed coupling $g_{utZ}=0.077$, see text for details) are imposed. The last column indicates the expected significance when both statistical and systematic uncertainties are taken into account.  }}

\label{Table3}
\end{table}

\subsubsection{THDMIII+DM}
As the last example, we discuss the monotop reach in the THDMIII+DM. The main difference between this model and the previous ones is that naturally the $h_2$ couplings to $t$ and $c$ dominate (instead of $u$). This leads to a large PDF suppression in the production cross-sections.  In addition, as explained in Sec.~\ref{discovery_reach}, in this case $\sigma(th_{2})\sim\sigma(\bar t h_2)$ and so no charge asymmetry is expected in the signal. As a consequence, the analysis should be done using both final state lepton's charges.

The cut-scanning for the signal is performed for the benchmark point $\tilde y^{tc}_u = 0.2$ (with all other $\tilde y$ entries put to zero) and $m_{h_{2}}=150 ~ \rm GeV$ (see the discussion in Sec.~\ref{sec:thdmiii} for details). We present in Table~\ref{Table4} the resulting cut-flow table when the following optimized variable cuts are applied:  one lepton (electron or muon) with $p_{T}> 20 ~\rm GeV$ and $|\eta|<2.5$, exactly one {\it b}-tagged jet with $p_{T}> 25 ~\rm GeV$ and $|\eta|<2.5$, $\slashed E_{T} > 80 ~ \rm GeV$, $M_{T} > 110 ~ \rm GeV$ and up to one jet with $25~{\rm GeV } < p_{T} < 90 ~\rm GeV$ and $|\eta|<4.5$. 
\begin{table}[h!t]
\renewcommand{\arraystretch}{1.5}
{\tiny
\begin{center}
\begin{tabular}{|c|c|c|c|c||c|c|c|c|c||c|c|}
\hline
{\it $\ell$} & {\it b-}jets &  $M_T$ & $\slashed E_{T}$& jets & $\sigma_{t\bar t j}$ & $\sigma_{tj}$ & $\sigma_{Wj}$  &$\sigma_{VV}$ & $\sigma_{signal}$& Sig. \\

($p_T > 20 ~\rm GeV$, & ($p_T > 25 ~\rm GeV$, & (GeV) & (GeV) & ($p_T < 90 ~\rm GeV$)& (pb) &(pb) &(pb)&(pb)& (pb) &  \\
$|\eta|<2.5$) &  $|\eta|<2.5$) & & & $|\eta|<4.5$) & & & & & &~ \\
\hline
- & - & - & - & - & 239 & 112.80 & 17035 & 84.20 & 0.32 &-\\ 

1 & - & - & - & - & 63.20 & 29.10  & 2740.90  & 17.70 & 0.04 & 0.13 \\

1 & 1 & - & - & - & 25.30  & 8.88 &  52.59 & 0.55 & 1.10*$10^{-2}$ & 0.19 \\

1 & 1 & $>$ 110 &- & - & 1.74 &  0.16 & 0.27  & 6.80*$10^{-3}$ & 5.40*$10^{-3}$ & 0.55 \\

1 & 1 & $>$ 110  & $>$ 80& - & 1.04 &  6.22*$10^{-2}$ & 1.80*$10^{-2}$ & 2.10*$10^{-3}$ & 4.30*$10^{-3}$ &  0.60 \\

1 & 1  & $>$ 110  & $>$ 80 & $\leq$ 1 & 0.40 & 4.20*$10^{-2}$ & 7.50*$10^{-3} $ & 1.93*$10^{-3}$ & 2.80*$10^{-3}$ & 0.62 \\

\hline
\end{tabular}
\end{center}
}
\caption{\footnotesize{Signal and background cross-sections before and after the proposed cuts for the THDMIII+DM are imposed. The signal is simulated for a reference point with $\tilde y^{tc}_u = 0.2$ and $m_{h_{2}}=150$ GeV. The last column indicates the expected significance when only statistical uncertainties are taken into account.}}

\label{Table4}
\end{table}
The key role of $M_T$ in differentiating signal from background is again clearly visible as well as the non correlation between $M_T$ and $\slashed E_{T}$. The final significance in this case is low, mainly due to the small signal cross-section. As a result, the THDMIII+DM is not likely to be probed with the monotop leptonic search strategy using the 2012 data. 

\begin{figure}[!h]
\begin{center}
\includegraphics[width=0.6 \textwidth]{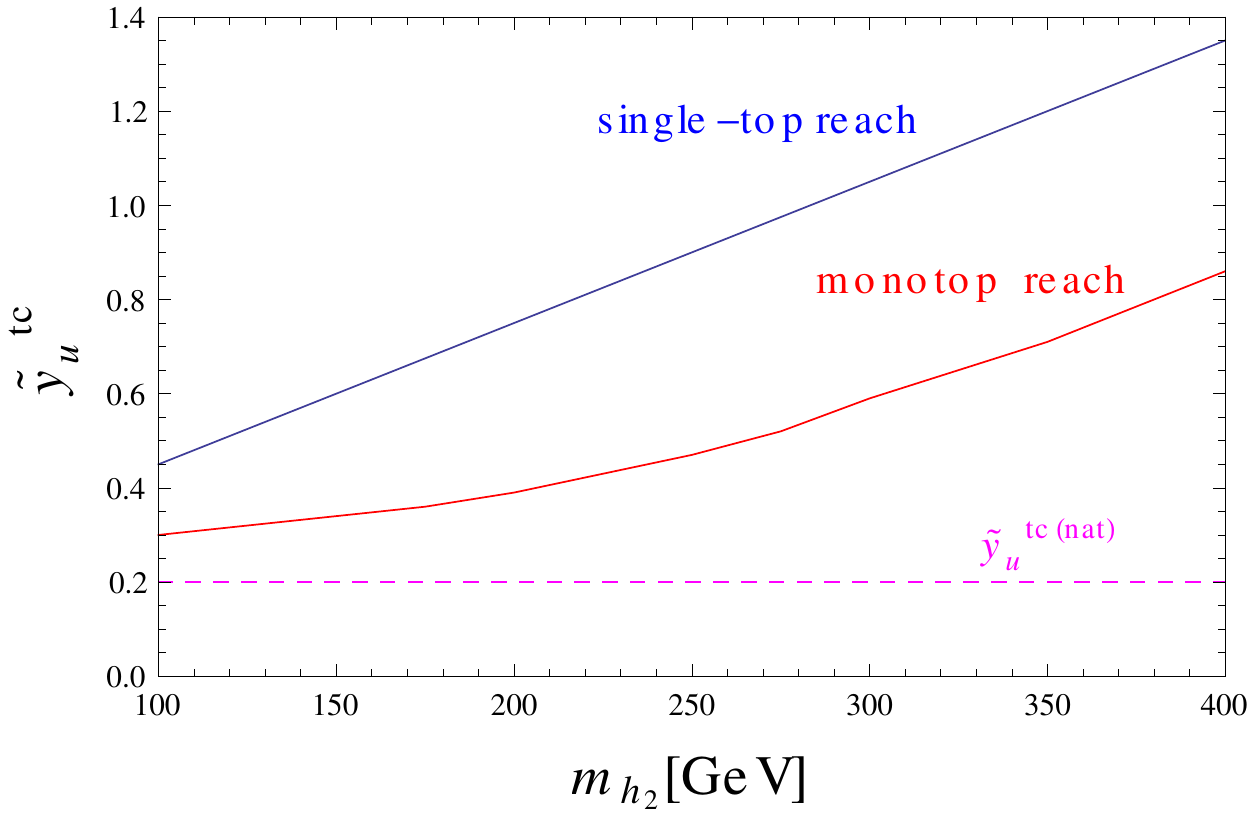}
\caption{\footnotesize  [Color online] The projected leptonic monotop and single top search strategies' reach with 2012 data in the  $\tilde y^{tc}_u -m_{h_{2}}$ plane for the THDMIII+DM. The single top search strategy corresponds to the one used in Ref.~\cite{1205.3130}. The natural value of $\tilde y^{tc}_u = 0.2$ as expected in the flavor model discussed in Sec.~\ref{sec:thdmiii} is marked by the horizontal  dashed magenta line.}
\label{monotop_single_reach_DM}
\end{center}
\end{figure}

Finally, we show in Fig.~\ref{monotop_single_reach_DM} the monotop and single top search strategies reaches in the $\tilde y_{u}^{ct}-m_{h_{2}}$ plane.  Note that, athough smaller than in the $Z'$ model case, there is a significant enhancement of the monotop leptonic reach over the single top one (Fig.~\ref{monotop_single_reach}). Also indicated in the plot is {the natural value of $\tilde y^{tc}_u = 0.2$ as expected in the flavor model discussed in Sec.~\ref{sec:thdmiii}}. It is clear that the proposed search strategy reach is not enough to probe such low values of $\tilde y^{tc}_u$ with existing available data.

\subsection{Discussion} \label{discussion}
Based on the monotop leptonic search strategies and their discovery/exclusion reach for the three example models studied in the previous subsection, we refer next to some general aspects of the presented analysis that we find  interesting to discuss in more detail.

We have investigated the background of the monotop leptonic signature and found that the main one comes from $t \bar t$ production, in contrast to the situation at the Tevatron, where $W j $ appears as the main background~\cite{Berger:1999zt,Berger:2000zk}. This is partly because the difference between $Wj$ and $t\bar t$ kinematic thresholds is less important at the LHC, and partly because $gg$ initial state contributions, which are more important in $t\bar t$ production, grow faster with collider energy. We have also verified that if our search strategy was applied at the Tevatron energies, then $W j$ would have effectively been the main background. 

Given that after all cuts our main background $t\bar t$ ends up being usually a couple of orders of magnitude bigger than $Wj$, it is worth recommending the experimental groups to consider the possibility of adjusting the $b$-tagging working point in order to reduce $t\bar t$ (and single top) at the price of increasing $Wj$.  In this work, we have employed the PGS original tune working point. If the $b$-tagging efficiency was increased -- at the price of increasing the contamination from light jets -- then the second $b$-jet in $t\bar t$ could be detected more efficiently, and those events could be discarded at event selection.  On the other hand, more $Wj$ events would pass the event selection because of increased fake b-tags.  Moreover, more signal events are expected to pass the $b$-jet requirement if the $b$-jet efficiency is increased.  The final balance should be an overall reduction in the background and an increase in the signal, yielding an increase in the final signal significance.
Finally, we note that this issue is expected to become even more important at larger LHC energies, since the dominant $t\bar t$ background is expected to become even more enhanced compared to other backgrounds and also the signal, and bringing it under control will become of utmost importance in order to further extend the reach of the leptonic monotop strategy.

We have also found the transverse mass of the lepton plus missing energy system, $M_T$, to be the most effective discriminator between the signal and backgrounds. We have explicitly shown in Tables~\ref{Table2} and~\ref{Table4} the effect of this variable cut on the simulated signal and background event samples, concluding that it is a key variable for this search strategy.  In particular, we have shown in Figs.~\ref{monotop_single_reach} and~\ref{monotop_single_reach_DM} that the monotop leptonic search strategy is significantly better than the single top one; we have seen that particularly for high masses of the invisible final state $X$ ($m_X\gtrsim 250$~GeV), the monotop search improves the existing single top one by up to an order of magnitude in the relevant coupling (or two orders of magnitude in the cross-section). As a matter of fact, although one could naively expect the single-top measurements to be sensitive to the monotop signature, this is not generally the case because for most single-top signatures within the SM, $M_T$ has an end point given by the $W$ mass.

We have also discussed the systematic uncertainties in the context of the FCNC $utZ$ interactions and found that they have little impact on the projected bound on $\mathcal B(t\to u Z)$. Moreover, in the case of the $Z'$ model and the FCNC $utZ$ interaction, where monotop production involves a valence $u$ quark in the initial state, we stress that an asymmetry in $\ell^\pm$ would suppress many systematic uncertainties, in addition to suppressing charge symmetric backgrounds such as $t\bar t$.  For instance, important systematic uncertainties may come from absolute cross-section measurements and luminosity uncertainties, which could represent an uncertainty of the order $\sim 15\%$ of the total number of events~\cite{experimentaltt}. To reduce these main systematic uncertainties which are common to positive ($N^+$) and negative ($N^-$) lepton events, one can construct an asymmetry such as $(N^+ - N^-)/(N^+ + N^-)$ and recover high levels of significance in the discovery/exclusion reach~\cite{1308.3712}. We note however, that this is  not applicable to the THDMIII+DM example, where monotops are expected to be predominantly produced via $c$ initial partons -- then the asymmetry is not present in the signal since $c$ and $\bar c$ PDFs coincide.

Finally, we would like to note that a combination of the proposed leptonic search strategy with a hadronic one~\cite{1106.6199,1308.3712,1109.5963} would allow to get a better discovery/exclusion reach for all considered scenarios. In addition, since this analysis was done assuming $21.7~ \rm fb^{-1}$ of luminosity, one should also expect a reach enhancement if CMS and ATLAS data were combined. The upcoming high energy/luminosity run of the LHC is naturally expected to further significantly extend this sensitivity.

\section{Conclusions}
\label{sec:5}

Monotops are predicted in many NP settings~\cite{1107.0623,Berger:1999zt,Berger:2000zk,Morrissey:2005uza,Dong:2011rh} and have lately invoked considerable theoretical interest~\cite{1106.6199,Berger:1999zt,Berger:2000zk,1308.3712,1109.5963}. In this work, we have performed a monotop leptonic search strategy motivated by the fact that this signature is cleaner than the hadronic mode and one can largely neglect the theoretically challenging QCD multijet background. We have investigated the discovery/exclusion reach with the LHC 2012 data in three different scenarios: a $Z'$ model that explains the apparent disagreement between the $A_{FB}$ and $A_{C}$ through a $utZ'$ interaction~\cite{1209.4354,1209.4872}, an effective $\Delta T=1$ FCNC $utZ$ interaction, and a THDMIII+DM example. 

We have studied and computed the existing constraints on each model and found that a stringent limit on the $Z'$ model can be derived from the existing (hadronic) monotop search at Tevatron~\cite{1202.5653}, while the single top production analysis by ATLAS ~\cite{1205.3130} does not constrain the relevant parameter space of the model. The larger reach of the CDF analysis in spite of the larger expected signal cross-section at ATLAS  is due to the fact that the experimental signal matches more closely the one predicted within the $Z'$ model. 
In the case of  FCNC $utZ$ interactions on the other hand, we have found that the latest CMS search for $t \to Zq$ decays imposes a far more stringent constraint on such contributions than existing single- and monotop analyses. Finally, neither indirect flavor constraints nor existing direct searches are yet sensitive to the interesting parameter space region in the THDMIII+DM example.

We have found that the main background of the leptonic monotop signature at the LHC is the SM $t \bar t$ production, while at the Tevatron the largest contribution comes from $W j$. Moreover, we have found that the transverse mass of the lepton plus missing energy system is the most powerful discriminating variable to distinguish signal from background. We have also compared the single top and monotop leptonic search strategies assuming existing 2012 LHC statistics. We have shown that dedicated searches for the monotop signature could allow to get substantially higher significances for all the models considered. These results are summarised in Figs.~\ref{monotop_single_reach} and~\ref{monotop_single_reach_DM}. 

While the $Z'$ model is already highly constrained, the remaining allowed parameter space could be completely covered using existing 2012 LHC dataset. In the case of FCNC ${utZ}$ interactions, we have instead found that the leptonic monotop search on its own cannot compete with the current sensitivity of FCNC top decay searches. This is mainly because of the lightness of the $Z$ boson in comparison to the top quark. Similarly,  the interesting parameter region of the THDMIII+DM will be difficult to probe with the 2012 data alone. We expect though that combining the leptonic and hadronic monotop search strategies on larger datasets expected from the high energy LHC run will allow to reach the relevant sensitivity also in such scenarios.  

We conclude that the proposed monotop search strategy is a promising new venue for discovering/excluding NP. We also stress that it can be applied to a variety of models predicting the monotop signature such as, for instance, R-parity violating supersymmetry~\cite{Berger:1999zt,Berger:2000zk} or baryon number violating interactions~\cite{Morrissey:2005uza,Dong:2011rh}, which have not been explicitly discussed in the present study.

\section*{Acknowledgments}
The authors are grateful to R. Piegaia and J. A. Aguilar-Saavedra for valuable discussions. E. C. L. would like to thank the J. Stefan Institute, where part of this work was done, for its hospitality.


\begin{thebibliography}{99}
\bibitem{topmass}
We use the reconstructed top mass definition in \cite{1205.3130}.
\end{thebibliography}
\end{document}